\documentclass[11pt]{article}
\usepackage{amssymb,amsmath,amsfonts}
\usepackage{graphicx}
\usepackage{graphics}
\usepackage{eepic,epsfig}
\usepackage{dcolumn}

1.60cm \makeatletter \@addtoreset{equation}{section}

\makeatother

\begin{document}

\title{Surface Casimir densities and induced cosmological  \\
constant on parallel branes in AdS}
\author{Aram A. Saharian\thanks{%
Email: saharyan@server.physdep.r.am} \\
\textit{Department of Physics, Yerevan State
University, 1 Alex Manoogian Str.} \\
\textit{ 375049 Yerevan, Armenia,}\\
\textit{and}\\
\textit{The Abdus Salam International Centre for Theoretical
Physics} \\
\textit{ 34014 Trieste, Italy }}
\date{\today}
\maketitle

\begin{abstract}
Vacuum expectation value of the surface energy-momentum tensor is
evaluated for a massive scalar field with general curvature
coupling parameter subject to Robin boundary conditions on two
parallel branes located on $(D+1)$-dimensional AdS bulk. The
general case of different Robin coefficients on separate branes is
considered. As an regularization procedure the generalized zeta
function technique is used, in combination with contour integral
representations. The surface energies on the branes are presented
in the form of the sums of single brane and second brane-induced
parts. For the geometry of a single brane both regions, on the
left (L-region) and on the right (R-region), of the brane are
considered. The surface densities for separate L- and R-regions
contain pole and finite contributions. For an infinitely thin
brane taking these regions together, in odd spatial dimensions the
pole parts cancel and the total surface energy is finite. The
parts in the surface densities generated by the presence of the
second brane are finite for all nonzero values of the interbrane
separation. It is shown that for large distances between the
branes the induced surface densities give rise to an exponentially
suppressed cosmological constant on the brane. In the
Randall-Sundrum braneworld model, for the interbrane distances
solving the hierarchy problem between the gravitational and
electroweak mass scales, the cosmological constant generated on
the visible brane is of the right order of magnitude with the
value suggested by the cosmological observations.
\end{abstract}

\bigskip

PACS number(s): 03.70.+k, 11.10.Kk, 04.62.+v

\bigskip

\section{Introduction}

\label{sec:introd}

The dynamics of fields on anti-de Sitter (AdS) background
possesses remarkable properties and much of early interest to this
spacetime was motivated by the questions of principal nature
related to the quantization of fields propagating on curved
backgrounds. The presence of the both regular and irregular modes
and the possibility of interesting causal structure lead to many
new phenomena. The importance of this theoretical work increased
when it has been discovered that AdS spacetime generically arises
as ground state in extended supergravity and in string theories.
Further interest in this subject was, and continues to be,
generated by the appearance of two models where AdS geometry plays
a special role. The first model, the so called AdS/CFT
correspondence, was motivated by Maldacena \cite{Mald98} (for a
review see \cite{Ahar00}) and represents a realization of the
holographic principle. The AdS/CFT correspondence relates string
theories or supergravity in the bulk of AdS with a conformal field
theory living on its boundary. It has many interesting formal and
physical facets and provides a powerful tool to investigate gauge
field theories, in particular QCD. The second model, suggested by
Randall and Sundrum \cite{Rand99a},\cite{Rand99b}, is a
realization of a braneworld scenario with large extra dimensions.
Recently it has been realized that the introduction of large extra
spatial dimensions may provide a solution to the hierarchy problem
between the gravitational and electroweak mass scales
\cite{Arka98} (for reviews in braneworld gravity and cosmology see
Refs. \cite{Ruba01}-\cite{Maar03}). The main idea to resolve the
large hierarchy is that the small coupling of four dimensional
gravity is generated by the large physical volume of extra
dimensions. Braneworlds naturally appear in string/M-theory
context and provide a novel setting for discussing
phenomenological and cosmological issues related to extra
dimensions. The model introduced by Randall and Sundrum is
particularly attractive. Their background solution consists of two
parallel flat branes, one with positive tension and another with
negative tension embedded in a five dimensional AdS bulk
\cite{Rand99a}. The fifth coordinate is compactified on $S^1/Z_2$,
and the branes are on the two fixed points. It is assumed that all
matter fields are confined on the branes and only the gravity
propagates freely in the five dimensional bulk. In this model, the
hierarchy problem is solved if the distance between the branes is
about 37 times the AdS radius and we live on the negative tension
brane. More recently, alternatives to confining particles on the
brane have been investigated and scenarios with additional bulk
fields have been considered \cite{Gher00},\cite{Poma00}. Apart
from the hierarchy problem it has been also tried to solve the
cosmological constant problem within the braneworld scenario. The
problem of the cosmological constant has been considered as the
most serious mass hierarchy problem in modern particle physics
(see, for instance, \cite{Wein89}) and many attempts addressing
this fine-tuning issue can be found in the literature. The
braneworld theories may give some alternative discussion of the
cosmological constant (see Refs. \cite{Arka00} and references
therein). The basic new ingredient is that the vacuum energy
generated by quantum fluctuations of fields living on the brane
may not curve the brane itself but instead the space transverse to
it.

The investigation of quantum effects in braneworld models is of
considerable phenomenological interest, both in particle physics
and in cosmology. The braneworld corresponds to a manifold with
dynamical boundaries and all fields which propagate in the bulk
will give Casimir-type contributions to the vacuum energy (for
reviews of the Casimir effect see Refs.
\cite{Most97},\cite{Milt04}), and as a result to the vacuum forces
acting on the branes. In dependence of the type of a field and
boundary conditions imposed, these forces can either stabilize or
destabilize the braneworld. In addition, the Casimir energy gives
a contribution to both the brane and bulk cosmological constants
and, hence, has to be taken into account in the self-consistent
formulation of the braneworld dynamics. Motivated by these, the
role of quantum effects in braneworld scenarios has received a
great deal of attention. For a conformally coupled scalar this
effect was initially studied in Ref. \cite{Fabi00} in the context
of M-theory, and subsequently in Refs.
\cite{Noji00a}-\cite{Saha04a} for a background Randall--Sundrum
geometry (for the related heat kernel expansions see Refs.
\cite{Bord99a}). The models with dS and AdS branes, and higher
dimensional brane models are considered as well
\cite{Eliz03},\cite{Noji00b}-\cite{Norm04}. For a conformally
cupled bulk scalar the cosmological backreaction of the Casimir
energy is investigated in Refs.
\cite{Fabi00},\cite{Eliz03},\cite{Noji00b},\cite{Muko01}-\cite{Yera03}.

In the papers mentioned above, the authors consider mainly the
global quantities such as the total Casimir energy or conformally
invariant fields. The investigation of local physical
characteristics in the Casimir effect, such as expectation value
of the energy-momentum tensor, is of considerable interest. In
addition to describing the physical structure of the quantum field
at a given point, the energy-momentum tensor acts as the source in
the Einstein equations and therefore plays an important role in
modelling a self-consistent dynamics involving the gravitational
field. In the case of two parallel branes on AdS background, the
vacuum expectation value of the bulk energy-momentum tensor for a
scalar field with an arbitrary curvature coupling is investigated
in Refs. \cite{Knap03},\cite{Saha04a}. In particular, in Ref.
\cite{Saha04a} the application of the generalized Abel-Plana
formula \cite{Sahreview} to the corresponding mode sums allowed us
to extract manifestly the parts due to the AdS spacetime without
boundaries and to present the boundary induced parts in terms of
exponentially convergent integrals for the points away the
boundaries. The interaction forces between the branes are
investigated as well. Depending on the coefficients in the
boundary conditions, these forces can be either attractive or
repulsive. On the background of manifolds with boundaries, the
physical quantities, in general, will receive both volume and
surface contributions and the surface terms play an important role
in various branches of physics. In particular, the surface
counterterms introduced to renormalize the divergencies in the
quasilocal definitions of the energy for the gravitational field
and in quantum field theory with boundaries are of particular
interest. For scalar fields with general curvature coupling, in
Ref. \cite{Rome02} it has been shown that in the discussion of the
relation between the mode sum energy, evaluated as the sum of the
zero-point energies for each normal mode of frequency, and the
volume integral of the renormalized energy density for the Robin
parallel plates geometry it is necessary to include in the energy
a surface term concentrated on the boundary (see also the
discussion in Refs. \cite{Milt04},\cite{Full03}). Similar issues
for the spherical and cylindrical boundary geometries are
discussed in Refs. \cite{Saha01},\cite{Rome01}. An expression for
the surface energy-momentum tensor for a scalar field with a
general curvature coupling parameter in the general case of bulk
and boundary geometries is derived in Ref. \cite{Saha03}. The
purpose of the present paper is to study the vacuum expectation
value of this tensor for a scalar field obeying Robin boundary
conditions on a single and two parallel branes of codimension one
in $(D+1)$-dimensional AdS spacetime. In particular, we show that
in the two-brane set up the surface densities induced on a brane
by the presence of the second brane are exponentially small for
large interbrane distances and give rise to naturally suppressed
cosmological constant in the brane universe.

The paper is organized  as follows. In the next section we show
that the vacuum expectation value of the surface energy momentum
tensor can be expressed via the expectation values of the field
square evaluated on the branes. By using the Cauchiy's theorem on
residues, an integral representations for the related zeta
functions for both branes are constructed. They contain parts due
to a single brane when the second brane is absent, and parts which
are induced by the presence of the second brane. The latters are
finite at the physical point. The analytic continuation for single
plate contributions is constructed in Section \ref{sec:1brane} for
both regions, on the right and on the left of the brane. The
surface densities for two-branes geometry are investigated in
Section \ref{sec:2brane}. Various limiting cases are considered
and the cosmological constant induced on the brane is estimated.
In Section \ref{sec:enbal} we discuss the balance between the
separate parts of the vacuum energy and show that they satisfy
standard thermodynamic relation. The last section contains a
summary of the work.

\section{Surface energy-momentum tensor and the generalized zeta function}

\label{sec:zetafunc}

In this paper we consider a massive scalar field $\varphi (x)$ on
background of a $(D+1)$-dimensional AdS spacetime ($AdS_{D+1}$)
with the line element
\begin{equation}
ds^{2}=g_{ik}dx^{i}dx^{k}=e^{-2k_{D}y}\eta _{\mu \nu }dx^{\mu }dx^{\nu
}-dy^{2},  \label{metric}
\end{equation}%
and AdS radius given by $1/k_{D}$. Here $\eta _{\mu \nu }={\rm diag}%
(1,-1,\ldots ,-1)$ is the metric for the $D$-dimensional Minkowski
spacetime, $i,k=0,1,\ldots ,D$, and $\mu ,\nu =0,1,\ldots ,D-1$.
The corresponding field equation reads
\begin{equation}
\left( g^{ik}\nabla _{i}\nabla _{k}+m^{2}+\zeta R\right) \varphi (x)=0,
\label{fieldeq}
\end{equation}%
where the symbol $\nabla _{i}$ is the operator for the covariant derivative
associated with the metric $g_{ik}$, $R=-D(D+1)k_{D}^{2}$ is the
corresponding Ricci scalar, and $\zeta $ is the curvature coupling
parameter. For minimally and conformally coupled scalars one has $\zeta =0$
and $\zeta =\zeta _{c}=(D-1)/4D$ correspondingly. Note that by making a
coordinate transformation
\begin{equation}
z=e^{k_{D}y}/k_{D},  \label{zcoord}
\end{equation}%
metric (\ref{metric}) is written in a manifestly conformally-flat form $%
ds^{2}=(k_{D}z)^{-2}\eta _{ik}dx^{i}dx^{k}$ with $x^D=z$. This is
the $AdS_{D+1}$ line element in Poincar\'{e} coordinates.

Below we will assume that the field obeys mixed boundary
conditions on two parallel infinite plane boundaries (branes),
located at $y=a$ and $y=b$, $a<b$:
\begin{equation}
\left( \tilde{A}_{y}+\tilde{B}_{y}\partial _{y}\right) \varphi
(x)=0,\quad y=a,b,  \label{boundcond}
\end{equation}%
with constant coefficients $\tilde{A}_{y}$, $\tilde{B}_{y}$. This
type of boundary conditions naturally arises for bulk fields in
the Randall-Sundrum braneworld due to the $Z_2$ symmetry of the
model. The presence of boundaries modifies the spectrum for the
zero--point fluctuations of the scalar field under consideration.
This leads to the modification of the vacuum expectation values of
physical quantities to compared with the case without boundaries.
For the geometry under consideration, the Wightman function and
the corresponding vacuum expectation value of the bulk
energy-momentum tensor are investigated in Ref. \cite{Saha04a}
(see also Ref. \cite{Saha03a} for the case of a conformally
invariant scalar field). The energy-momentum tensor for a scalar
field on manifolds with boundaries in addition to the bulk part
contains a contribution located on the boundary. For an arbitrary
smooth boundary $\partial M_{s}$ with the inward-pointing unit
normal vector $n^l$, the surface part of the energy-momentum
tensor is given by the formula \cite{Saha03}
\begin{equation}
T_{ik}^{{\rm (surf)}}=\delta (x;\partial M_{s})\tau _{ik}  \label{Ttausurf}
\end{equation}%
with
\begin{equation}
\tau _{ik}=\zeta \varphi ^{2}K_{ik}-(2\zeta -1/2)h_{ik}\varphi n^{l}\nabla
_{l}\varphi ,  \label{tausurf}
\end{equation}%
and the "one-sided" delta-function $\delta (x;\partial M_{s})$
locates this tensor on $\partial M_{s}$. In Eq. (\ref{tausurf}),
$K_{ik}=h_i^l h_k^m \nabla _ln_m$ is the extrinsic curvature
tensor of the boundary $\partial M_{s}$ and $h_{ik}=g_{ik}+n_in_k$
is the corresponding induced metric. Here we are interested in the
vacuum expectation values of the surface energy-momentum tensor
(\ref{Ttausurf}) on the branes $y=a$ and $y=b$.

Let $\{\varphi _{\alpha }(x),\varphi _{\alpha }^{\ast }(x)\}$ be a
complete set of positive and negative frequency solutions to the
field equation (\ref{fieldeq}), obeying the boundary condition
(\ref{boundcond}). Here $\alpha $ is a collective index for all
quantum numbers. By expanding the field operator over the
eigenfunctions $\varphi _{\alpha }(x)$, using the standard
commutation rules and the definition of the vacuum state, for the
vacuum expectation value of the surface energy-momentum tensor one
finds
\begin{equation}
\langle 0|T_{ik}^{{\rm (surf)}}|0\rangle =\delta (x;\partial
M_{s})\langle 0|\tau _{ik}|0\rangle ,\label{modesumform1}
\end{equation}%
where
\begin{equation}\label{modesumform}
\langle 0|\tau _{ik}|0\rangle =\sum_{\alpha }\tau _{ik}\{\varphi
_{\alpha }(x),\varphi _{\alpha }^{\ast }(x)\}.
\end{equation}
Here $|0\rangle $ is the amplitude for the corresponding vacuum
state, and the bilinear form $\tau _{ik}\{\varphi ,\psi \}$ on the
right of the second formula is determined by the classical
energy-momentum tensor (\ref{tausurf}). For the geometry of two
parallel branes on AdS bulk one has ($j=a,b$)
\begin{equation}
\begin{split}
n^{(j)l} &=n^{(j)}\delta _{D}^{l},\quad n^{(a)}=1,\quad
n^{(b)}=-1,
\label{njKj} \\
K_{\mu \nu }^{(j)} &=-n^{(j)}k_{D}g_{\mu \nu },\quad \mu ,\nu
=0,1,\ldots ,D-1,
\end{split}
\end{equation}
and $K_{DD}^{(j)}=0$, where $n^{(j)l}$ and $K_{ik}^{(j)}$ are the
inward-pointing unit normal and the extrinsic curvature tensor for
the brane at $y=j$, $j=a,b$ (we consider the region between the
branes, $a\leq y\leq b$). Note that boundary conditions
(\ref{boundcond}) can be written in the covariant form $( \tilde
A_j+n^{(j)}\tilde B_j n^{(j)l}\nabla _l ) \varphi =0$. By using
relations (\ref{njKj}) and the boundary conditions, the vacuum
expectation value of the surface energy-momentum tensor on the
brane at $y=j$ is presented in the form
\begin{equation}
\langle 0|\tau _{\mu \nu }^{(j)}|0\rangle =-g_{\mu \nu }n^{(j)}
\left[ \zeta k_{D}-(2\zeta -1/2)\tilde{A}_{j}/\tilde{B}_{j}\right]
\langle 0|\varphi ^{2}|0\rangle _{z=z_j},  \label{tauj}
\end{equation}
and $\langle 0|\tau _{DD}^{(j)}|0\rangle =0$. From the point of
view of physics on the brane, Eq. (\ref{tauj}) corresponds to the
gravitational source of the cosmological constant type,
\begin{equation}\label{tauj1}
\langle 0|\tau _{\mu }^{(j)\nu }|0\rangle ={\mathrm{diag}} \left(
\varepsilon ^{{\mathrm{(surf)}}}_j,-p^{{\mathrm{(surf)}}}_j,\ldots
, -p^{{\mathrm{(surf)}}}_j \right) ,
\end{equation}
with the surface energy density $\varepsilon
^{{\mathrm{(surf)}}}_j$ (surface energy per unit physical volume
on the brane at $z=z_j$ or brane tension), stress
$p^{{\mathrm{(surf)}}}_j$ and the equation of state
\begin{equation}\label{eqstate}
\varepsilon ^{{\mathrm{(surf)}}}_j=-p^{{\mathrm{(surf)}}}_j .
\end{equation}
Of course, this is the direct consequence of the Poincar\'{e}
invariance of the branes.

For an untwisted bulk scalar in the $(D+1)$-dimensional version of
the Randall-Sundrum braneworld with brane mass terms $c_a$ and
$c_b$, the ratio of the coefficients in the boundary condition
(\ref{boundcond}) is determined by the expression (see, e.g.,
Refs. \cite{Gher00},\cite{Flac01b},\cite{Saha04a})
\begin{equation}\label{AjBjRS}
    \frac{\tilde A_j}{\tilde B_j}=-\frac{n^{(j)}c_j+4D\zeta k_D}{2}.
\end{equation}
For the coefficient in formula (\ref{tauj}) this gives
\begin{equation}\label{gorcRS}
2\zeta -(4\zeta -1)\frac{\tilde A_j}{k_D\tilde B_j}=8D\zeta (\zeta
-\zeta _c)+\left( 4\zeta -1\right) n^{(j)} \frac{c_j}{2k_D} .
\end{equation}
In particular, the surface energy in the Randall-Sundrum
braneworld vanishes for minimally and conformally coupled scalar
fields with zero brane mass terms. Note that in the supersymmetric
version of the model \cite{Gher00} one has $c_b=-c_a$.

AdS spacetime is divided by the branes into three regions with
$y\leq a$, $a\leq y\leq b$, and $y\geq b$. Below in this section,
we will consider the region between the branes. The corresponding
quantities for the other regions are obtained from those as
limiting cases and are investigated in the next section. Note that
in a $S^{1}/Z_2$ version of the model, the only bulk is between
the branes. Due to the $D$-dimensional Poincar\'{e} invariance,
the corresponding eigenfunctions can be presented in the form
\begin{equation}
\begin{split}
\varphi _{\alpha }(x)&=\frac{f_{u}(y)e^{-i\eta _{\mu \nu }k^{\mu }x^{\nu }}}{%
\sqrt{2\omega (2\pi )^{D-1}}},\quad k^{\mu }=(\omega ,{\bf k}),
\label{eigfunc1} \\
\omega &= \sqrt{k^{2}+u^{2}},\quad k=|{\bf k}|.
\end{split}
\end{equation}%
Here the separation constants $u$ are determined by boundary conditions (\ref%
{boundcond}) and will be given below. Substituting eigenfunctions (\ref%
{eigfunc1}) into the field equation (\ref{fieldeq}), one obtains
the equation for the function $f_{u}(y)$ with the solution
\begin{equation}
f_{u}(y)=C_{\alpha }e^{Dk_{D}y/2}\left[ J_{\nu }(uz)+b_{\nu
}Y_{\nu }(uz)\right] , \label{fny}
\end{equation}%
in the region $a\leq y\leq b$. Here $J_{\nu }(x)$, $Y_{\nu }(x)$
are the Bessel and Neumann functions of the order
\begin{equation}
\nu =\sqrt{(D/2)^{2}-D(D+1)\zeta +m^{2}/k_{D}^{2}}.  \label{nu}
\end{equation}%
The parameter $\nu $ must be real to ensure stability
\cite{Brei82},\cite{Mezi85}. For a given $\zeta $ this imposes a
lower bound for the mass (Breitenlohner-Freedman bound). Note that
on AdS bulk the parameter $m^2$ can be negative. In the case of a
conformally coupled massless scalar, $\zeta =\zeta _c$, one has
$\nu =1/2$ and the cylinder functions in Eq. (\ref{fny}) are
expressed via the elementary functions.

In the region between the branes, the coefficient $b_{\nu }$ is
determined from the boundary condition on $y=a$:
\begin{equation}
b_{\nu }=-\frac{\bar{J}_{\nu }^{(a)}(uz_{a})}{\bar{Y}_{\nu }^{(a)}(uz_{a})}%
,\quad z_{j}=\frac{e^{k_{D}j}}{k_{D}},\quad j=a,b\, . \label{benu}
\end{equation}%
Here and below we use the barred notation
\begin{equation}
\bar{F}^{(j)}(x)=A_{j}F(x)+B_{j}xF^{\prime }(x), \label{notbar}
\end{equation}%
for a given function $F(x)$, with the coefficients
\begin{equation}\label{AAtilda}
A_{j}=\tilde{A}_{j}+%
\tilde{B}_{j}k_{D}D/2,\quad B_{j}=\tilde{B}_{j}k_{D}.
\end{equation}
Note that for a bulk scalar field in the Randall-Sundrum model
from Eq. (\ref{AjBjRS}) one has $2A_j/B_j=D(1-4\zeta
)-n^{(j)}c_{j}/k_D$. From the boundary condition on the brane
$y=b$ we receive that the eigenvalues for $u$ have to be solutions
to the equation
\begin{equation}
g_{\nu }^{(ab)}(uz_{a},uz_{b})\equiv \bar{J}_{\nu }^{(a)}(uz_{a})\bar{Y}%
_{\nu }^{(b)}(uz_{b})-\bar{Y}_{\nu }^{(a)}(uz_{a})\bar{J}_{\nu
}^{(b)}(uz_{b})=0.  \label{cnu}
\end{equation}%
This equation gives the spectrum of Kaluza-Klein masses.
We denote by $u=u_{\nu ,n}$, $n=1,2,\ldots $, the zeros of the function $%
g_{\nu }^{(ab)}(uz_{a},uz_{b})$ in the right half-plane of the
complex variable $u$, arranged in the ascending order, $u_{\nu
,n}<u_{\nu ,n+1}$. The set of quantum numbers specifying the
eigenfunctions is $\alpha =({\mathbf{k}},n)$. In our analysis we
will assume that the values $A_j/B_j$ are such that there are no
imaginary zeros. This is the case, for example, when $A_a/B_a\leq
0$ and $A_b/B_b\geq 0$ (see also Ref. \cite{Saha03mon}). The
coefficient $C_{\alpha }$ in Eq. (\ref{fny}) is determined from
the orthonormality condition
\begin{equation}
\int_{a}^{b}dy\, e^{(2-D)k_{D}y}f_{u_{\nu ,n}}(y)f_{u_{\nu
,n^{\prime }}}(y)=\delta _{nn^{\prime }} , \label{ortcond}
\end{equation}%
and is equal to
\begin{equation}
C^{2}_{\alpha }=\frac{\pi }{k_{D}z_{a}}\frac{\bar{Y}_{\nu
}^{(a)}(uz_{a})\bar{Y}_{\nu
}^{(b)}(uz_{b})}{\frac{\partial }{\partial u}g_{\nu }^{(ab)}(uz_{a},uz_{b})}%
,\quad u=u_{\nu ,n}.  \label{cn}
\end{equation}%

As it follows from formula (\ref{tauj}), the vacuum expectation
values of the surface energy-momentum tensor can be obtained from
the vacuum expectation values of the field square evaluated on the
branes. Substituting the eigenfunctions (\ref{eigfunc1}) into the
corresponding mode sum and integrating over the angular part of
the vector ${\bf k}$, for the expectation value of the field
square in the region between the branes one finds
\begin{eqnarray}
\langle 0|\varphi ^{2}(x)|0\rangle &=&\sum_{\alpha }\varphi
_{\alpha
}(x)\varphi _{\alpha }^{\ast }(x)  \nonumber \\
&=&\pi k_{D}^{D-1}z^{D}\beta _{D} \int_{0}^{\infty }dk\,\sum_{n=1}^{\infty }\frac{%
k^{D-2}u_{\nu ,n}\prod_{j=a,b}g_{\nu }^{(j)}(u_{\nu ,n}z_{j},u_{\nu ,n}z)}{\sqrt{%
u_{\nu ,n}^{2}+k^{2}}\left. \frac{\partial }{\partial u}g_{\nu
}^{(ab)}(uz_{a},uz_{b})\right| _{u=u_{\nu ,n}}},  \label{W11}
\end{eqnarray}%
where we have introduced the notations%
\begin{equation}\label{betaD}
\beta _{D}=\frac{1}{(4\pi )^{\frac{D-1}{2}}\Gamma \left( \frac{D-1%
}{2}\right) } ,
\end{equation}
and
\begin{equation}
g_{\nu }^{(j)}(u,v)=J_{\nu }(v)\bar{Y}_{\nu }^{(j)}(u)-Y_{\nu }(v)\bar{J}%
_{\nu }^{(j)}(u).  \label{gnuj}
\end{equation}%
By using the relation $g_{\nu }^{(j)}(u,u)=2B_{j}/\pi $, for the
corresponding vacuum expectation value on the brane at $y=j$ we
obtain
\begin{equation}
\langle 0|\varphi ^{2}(x)|0\rangle _{z=z_{j}}=
2k_{D}^{D-1}z_{j}^{D}B_{j}\beta _{D} \int_{0}^{\infty
}dk\,k^{D-2}\sum_{n=1}^{\infty }\frac{u_{\nu ,n}g_{\nu
}^{(l)}(u_{\nu ,n}z_{l},u_{\nu ,n}z_{j})}{\sqrt{u_{\nu
,n}^{2}+k^{2}}\left. \frac{\partial }{\partial u}g_{\nu
}^{(ab)}(uz_{a},uz_{b})\right| _{u=u_{\nu ,n}}},  \label{phi2j}
\end{equation}%
where $j,l=a,b$, and $l\neq j$. Quantity (\ref{phi2j}) and, hence,
the surface energy-momentum tensor diverge and need some
regularization. Many regularization techniques are available
nowadays and, depending on the specific physical problem under
consideration, one of them may be more suitable than the others.
In particular, the generalized zeta function method
\cite{Dowk76},\cite{Eliz94} is in general very powerful to give
physical meaning to the divergent quantities. There are several
examples of the application of this method to the evaluation of
the Casimir effect (see, for instance, \cite{Eliz94}-\cite{Blau88}
and references therein). Here we will use the method which is an
analog of the generalized zeta function approach.

Instead of (\ref{phi2j}) we define the function
\begin{equation}
F_{j}(s)=2k_{D}^{D-1}z_{j}^{D}\frac{B_{j}\beta _{D}}{\mu ^{1+s}}
\int_{0}^{\infty }dk\,k^{D-2}\sum_{n=1}^{\infty }(u_{\nu
,n}^{2}+k^{2})^{s/2}\frac{u_{\nu
,n}g_{\nu }^{(l)}(u_{\nu ,n}z_{l},u_{\nu ,n}z_{j})}{\frac{\partial }{%
\partial u}g_{\nu }^{(ab)}(uz_{a},uz_{b})|_{u=u_{\nu ,n}}},  \label{IAs}
\end{equation}%
with $\mu $ an arbitrary mass scale which has been introduced to
keep the dimension of the expression. Evaluating the integral
over $k$, this expression can be presented in the form%
\begin{equation}
F_{j}(s)=k_{D}^{D-1}z_{j}^{D}\frac{B_{j}\beta _{D}}{\mu ^{1+s}}
B\left( \frac{D-1}{2},-\frac{D-1+s}{2} \right) \zeta _{j}(s),
\label{Fjs2}
\end{equation}%
where $B(x,y)$ is the beta function and we have defined the
generalized zeta function as
\begin{equation}
\zeta _{j}(s)=\sum_{n=1}^{\infty }\frac{u_{\nu ,n}^{D+s}g_{\nu
}^{(l)}(u_{\nu ,n}z_{l},u_{\nu ,n}z_{j})}{\frac{\partial
}{\partial u}g_{\nu }^{(ab)}(uz_{a},uz_{b})|_{u=u_{\nu ,n}}}.
\label{zetsx}
\end{equation}%
We could include in the definition of the zeta function an
additional factor $\mu ^{-D-s-2}$ to keep this function
dimensionless. However, this will not affect on the final result
for the analytic continuation of the function $F_{j}(s)$. The
computation of vacuum expectation value of the surface
energy-momentum tensor requires the analytic continuation of the
function $F_{j}(s)$ to the value $s=-1$ (here and below $|_{s=-1}$
is understood in the sense of the analytic continuation),
\begin{equation}
\langle 0|\varphi ^{2}|0\rangle _{z=z_{j}}=F_{j}(s)|_{s=-1}.  \label{IFs0}
\end{equation}

The starting point of our consideration is the representation of the
function (\ref{zetsx}) in terms of contour integral:
\begin{equation}
\zeta _{j}(s)=\frac{1}{2\pi i}\int_{C}du\,u^{D+s}\frac{g_{\nu
}^{(l)}(uz_{l},uz_{j})}{g_{\nu }^{(ab)}(uz_{a},uz_{b})},  \label{intzetsx1}
\end{equation}%
where $C$ is a closed counterclockwise contour in the complex $u$
plane enclosing all zeros $u_{\nu ,n}$. The location of these
zeros \ enables one to deform the contour $C$ into a segment of
the imaginary axis $(-iR,iR)$ and a semicircle of radius $R$,
$R\to \infty $, in the right half-plane. We will also assume that
the origin is avoided by the semicircle $C_{\rho }$ in the right
half-plane with small radius $\rho $. For sufficiently large $s$
the integral over the large semicircle in Eq. (\ref{intzetsx1})
tends to zero in the limit $R\rightarrow \infty $, and the
expression on the right can be transformed to
\begin{equation}
\begin{split}
\zeta _{j}(s) = & \frac{1}{2\pi i}\int_{C_{\rho
}}du\,u^{D+s}\frac{g_{\nu
}^{(l)}(uz_{l},uz_{j})}{g_{\nu }^{(ab)}(uz_{a},uz_{b})}  \\
&+\frac{1}{\pi }\sin \frac{\pi }{2}\left( D+1+s\right) \int_{\rho
}^{\infty }du\,u^{D+s}\frac{G_{\nu }^{(l)}(uz_{l},uz_{j})}{G_{\nu
}^{(ab)}(uz_{a},uz_{b})},  \label{intzetsx2}
\end{split}
\end{equation}%
where ($j=a,b$)
\begin{eqnarray}
G_{\nu }^{(j)}(u,v) &=&I_{\nu }(v)\bar{K}_{\nu }^{(j)}(u)-K_{\nu }(v)\bar{I}%
_{\nu }^{(j)}(u),  \label{Gnuj12} \\
G_{\nu }^{(ab)}(u,v) &= & \bar{K}_{\nu }^{(a)}(u)\bar{I}_{\nu }^{(b)}(v)-%
\bar{I}_{\nu }^{(a)}(u)\bar{K}_{\nu }^{(b)}(v),  \label{Gnuab}
\end{eqnarray}%
and we have introduced the Bessel modified functions $I_{\nu }(u)$
and $K_{\nu }(u)$. Below we will consider the limit $\rho
\rightarrow 0$. In this limit the first integral on the right of
formula (\ref{intzetsx2}) vanishes at the physical point $s=-1$,
and we will concentrate on the contribution of the second
integral. The corresponding expression can be presented in the
form
\begin{equation}\label{zet12}
\zeta _{j}(s) =-\frac{1}{\pi }\sin \frac{\pi }{2}\left(
D+1+s\right) \int_{\rho }^{\infty }du\,u^{D+s}\left[
n^{(j)}U_{j\nu }(uz_j)+B_{j}\Omega _{j\nu }(uz_{a},uz_{b})\right]
,
\end{equation}
with the notations%
\begin{equation}\label{Unuj}
U_{a\nu }(x)=\frac{K_{\nu }(x)}{\bar{K}_{\nu
}^{(a)}(x)},\quad U_{b\nu }(x)=\frac{I_{\nu }(x)}{\bar{I}%
_{\nu }^{(b)}(x)}
\end{equation}
and
\begin{equation}\label{Omegaab}
\begin{split}
\Omega _{a\nu }(u,v) & =\frac{\bar{K}_{\nu
}^{(b)}(v)}{\bar{K}_{\nu
}^{(a)}(u) G_{\nu }^{(ab)}(u,v)}, \\
\Omega _{b\nu }(u,v) & =\frac{\bar{I}_{\nu
}^{(a)}(u)}{\bar{I}_{\nu }^{(b)}(v) G_{\nu }^{(ab)}(u,v)}.
\end{split}
\end{equation}%
Now by using the formula%
\begin{equation}
B\left( \frac{D-1}{2},-\frac{D-1+s}{2}\right) \sin \frac{\pi }{2}\left(
D+1+s\right) =\frac{\pi \Gamma \left( \frac{D-1}{2}\right) }{\Gamma \left( -%
\frac{s}{2}\right) \Gamma \left( \frac{D+1+s}{2}\right) },  \label{BetGam}
\end{equation}%
the corresponding expressions for the functions $F_{j}(s)$ can be
rewritten in the form%
\begin{equation}
F_{j}(s) =-\frac{(4\pi
)^{\frac{1-D}{2}}k_{D}^{D-1}z_{j}^{D}B_{j}}{\Gamma \left(
-\frac{s}{2}\right) \Gamma \left( \frac{D+1+s}{2} \right) \mu
^{s+1}}\int_{\rho }^{\infty }du\,u^{D+s}\left[ n^{(j)}U_{j\nu
}(uz_j)+B_{j}\Omega _{j\nu }(uz_{a},uz_{b})\right]  . \label{Fab}
\end{equation}%
The contribution of the second term in the square brackets is
finite at $s=-1$ and vanishes in the limits $z_a\to 0$ and $z_b\to
\infty $. The first term in the square brackets of this expression
corresponds to the contribution of a single brane at $z=z_j$ when
the second brane is absent. The regularization is needed for this
term only and the corresponding analytic continuation has been
done in the next section.

\section{Surface energy-momentum tensor for a single brane}

\label{sec:1brane}

The single brane at $z=z_{a}$ divides the AdS spacetime into two
regions corresponding to $z<z_{a}$ (L-region) and $z>z_{a}$
(R-region). The properties of the vacuum in these regions are
different and the corresponding quantities we will differ by the
indices R (right) and L (left), respectively. Let us consider
these regions separately.

\subsection{L-region}

\label{subsec:Lregion}

From Eq. (\ref{Fab}) it follows that for the geometry of a single
brane located at $z=z_{a}$ the function $F_{a}(s)\equiv
F_{a}^{{\rm (L)}}(s)$ is determined by the expression
\begin{equation}
F_{a}^{{\rm (L)}}(s)=\frac{(4\pi )^{\frac{1-D}{2}}k_{D}^{D-1}
B_{a}}{\Gamma \left( -\frac{s}{2}\right) \Gamma \left(
\frac{D+1+s}{2}\right) (\mu z_{a})^{1+s}}\int_{0}^{\infty
}du\,u^{D+s}\frac{I_{\nu }(u)}{\bar{I}_{\nu }^{(a)}(u)}.
\label{FLs}
\end{equation}%
This integral representation is valid in the strip $-(D+1)<{\rm
Re}\,s<-D$ and under the assumption that the function
$\bar{I}_{\nu }^{(a)}(u)$ has no real zeros. The latter
corresponds to the absence of imaginary zeros for the
eigenfunction $f_u(y)$ in Eq. (\ref{eigfunc1}) with respect to
$u$. This condition is satisfied for $A_a/B_a\geq -\nu $. For the
analytic continuation to $s=-1$, we write the integral on the
right of this formula as the sum of the integrals over the
intervals $(0,1)$ and $(1,\infty )$. The first integral is finite
at $s=-1$. To find an analytic continuation of the second integral
we employ the asymptotic expansions of the Bessel modified
function for large values of the argument (see, for instance,
\cite{abramowiz}). For $ B_{a}\neq 0$ from these expansions  one
has
\begin{equation}
\frac{I_{\nu }(u)}{\bar{I}_{\nu }^{(a)}(u)}\sim \frac{1}{B_{a}}%
\sum_{l=0}^{\infty }\frac{w_{l}(\nu )}{u^{l+1}},  \label{Inuratas}
\end{equation}%
where the coefficients $w_{l}(\nu )$ are combinations of the
corresponding coefficients in the expansions for the functions
$I_{\nu }(u)$ and $I_{\nu }^{\prime }(u)$. The first four
coefficients are as follows:
\begin{equation}
\begin{split}
w_{0}(\nu )&=1,\quad w_{1}(\nu )=\frac{1}{2}-\frac{A_{a}}{B_{a}},
\\
w_{2}(\nu )&=\frac{3}{8}-\frac{A_{a}}{B_{a}}+
\frac{A_{a}^{2}}{B_{a}^{2}}
-\frac{\nu ^{2}}{2},  \\
w_{3}(\nu )&=\frac{3}{8}+\frac{3}{2} \frac{A_{a}^{2}}{B_{a}^{2}}
-\frac{A_{a}^{3}}{B_{a}^{3}}+\frac{A_{a}}{B_{a}}\left( \nu
^{2}-1\right) -\nu ^{2}. \label{sphw}
\end{split}
\end{equation}%
Now we subtract and add to the integrand of the integral over
$(1,\infty )$ the $N$ leading terms of the corresponding
asymptotic expansion and exactly integrate the asymptotic part. By
this way Eq. (\ref{FLs}) may be written in the form
\begin{equation}
\begin{split}
F_{a}^{{\rm (L)}}(s) = & \frac{(4\pi
)^{\frac{1-D}{2}}k_{D}^{D-1}(\mu
z_{a})^{-1-s}}{\Gamma \left( -\frac{s}{2}\right) \Gamma \left( \frac{D+1+s}{2%
}\right) }\left\{ B_{a}\int_{0}^{1}du\,u^{D+s}\frac{I_{\nu }(u)}{\bar{I}%
_{\nu }^{(a)}(u)}\right.  \\
&+\left. \int_{1}^{\infty }du\,u^{D+s}\left[ B_{a}\frac{I_{\nu }(u)}{\bar{I}%
_{\nu }^{(a)}(u)}-\sum_{l=0}^{N}\frac{w_{l}(\nu )}{u^{l+1}}\right]
-\sum_{l=0}^{N}\frac{w_{l}(\nu )}{D+s-l}\right\} .  \label{FLs1}
\end{split}
\end{equation}%
For $N\geq D-1$ both integrals on the right are finite at $s=-1$ and $F_{a}^{%
{\rm (L)}}(s)$ has a simple pole at $s=-1$ corresponding to the summand $%
l=D-1$ in the last sum. As a result, the expression for the vacuum
expectation value for the field square on the brane contains pole
and finite parts:
\begin{equation}
\langle 0|\varphi ^{2}|0\rangle _{z=z_{a}}^{{\rm (L)}}=\langle \varphi
^{2}\rangle _{z=z_{a},{\rm p}}^{{\rm (L)}}+\langle \varphi ^{2}\rangle
_{z=z_{a},{\rm f}}^{{\rm (L)}}.  \label{phi2Lpf}
\end{equation}%
Laurent-expanding the expression on the right of Eq. (\ref{FLs1})
near $s=-1$, one finds%
\begin{equation}
\langle \varphi ^{2}\rangle _{z=z_{a},{\rm p}}^{{\rm (L)}}
=-2k_{D}^{D-1}w_{D-1}(\nu )\frac{\beta _{D+1}}{s+1}, \label{FL-1}
\end{equation}
for the pole part, and
\begin{equation}
\begin{split}
\langle \varphi ^{2}\rangle _{z=z_{a},{\rm f}}^{{\rm (L)}} = &
2k_{D}^{D-1}\beta _{D+1} \left\{
B_{a}\int_{0}^{1}du\,u^{D-1}\frac{I_{\nu }(u)}{\bar{I}_{\nu
}^{(a)}(u)}\right.  \\
&+\int_{1}^{\infty }du\,u^{D-1}\left[ B_{a}\frac{I_{\nu
}(u)}{\bar{I}_{\nu }^{(a)}(u)}-\sum_{l=0}^{N}\frac{w_{l}(\nu
)}{u^{l+1}}\right]
-\sum_{\substack{l=0\\ l\neq D-1}}^{N}\frac{w_{l}(\nu )}{D-l-1}  \\
&+\left. w_{D-1}(\nu )\left[ \ln (\mu z_{a})+\frac{1}{2}\psi \left( \frac{D%
}{2}\right) -\frac{1}{2}\psi \left( \frac{1}{2}\right) \right] \right\} ,
\label{phi2Lpf2}
\end{split}
\end{equation}%
for the finite part, with $\psi (x)$ being the diagamma function.
By using relations (\ref{tauj}) and (\ref{phi2Lpf}), we can write
analogous decomposition for the surface energy density in the
L-region:
\begin{eqnarray}
&& \varepsilon _{a}^{{\rm (surf)(L)}}=\varepsilon _{a,{\rm
p}}^{{\rm (surf)(L)}}+\varepsilon _{
a,{\rm f}}^{{\rm (surf)(L)}}, \label{epspf} \\
&& \varepsilon _{a,{\rm s}}^{{\rm (surf)(L)}}=\left[ \zeta
k_{D}-(2\zeta -1/2)\tilde{A}_{a}/\tilde{B}_{a}\right] \langle
\varphi ^{2}\rangle _{z=z_{a},{\rm s}}^{{\rm (L)}},
\end{eqnarray}%
with ${\rm s=p,f}$, and for the surface stresses via the equation
of state. Note that in the principal part prescription adopted in
Ref. \cite{Blau88} (see also \cite{Eliz94}) the Casimir energy
corresponds to the principal part of the corresponding Laurent
expansion near the physical point. In our case this is presented
by the quantity $\varepsilon _{a,{\rm f}}^{{\rm (surf)(L)}}$ (
$\varepsilon _{ a,{\rm f}}^{{\rm (surf)(R)}}$ for the R-region,
see below).

\subsection{R-region}

For a single brane at $z=z_{a}$ the function $F_{a}(s)\equiv F_{a}^{{\rm (R)}%
}(s)$ in the R-region is determined from Eq. (\ref{Fab}):
\begin{equation}
F_{a}^{{\rm (R)}}(s)=-\frac{(4\pi )^{\frac{1-D}{2}}k_{D}^{D-1}
B_{a}}{\Gamma \left( -\frac{s}{2}\right) \Gamma \left(
\frac{D+1+s}{2}\right) (\mu z_{a})^{1+s}}\int_{0}^{\infty
}du\,u^{D+s}\frac{K_{\nu }(u)}{\bar{K}_{\nu }^{(a)}(u)},
\label{FRs}
\end{equation}
and differs from the corresponding function in the L-region by the
replacement $I_{\nu }\to K_{\nu }$ (in general, the coefficients
in the boundary condition could be different for the left and
right surfaces of the brane). This integral representation is
valid in the strip $-(D+1)<{\rm Re}\,s<-D$ and under the
assumption of absence of real zeros for the function $\bar{K}_{\nu
}^{(a)}(u)$. The latter is the case for $A_a/B_a\leq \nu $. The
analytic continuation of the integral on the right to $s=-1$ can
be found by the way similar to that for the L-region. For $
B_{a}\neq 0$ the corresponding asymptotic expansion of the
subintegrand for large values of the argument has the form
\begin{equation}
\frac{K_{\nu }(u)}{\bar{K}_{\nu }^{(a)}(u)}\sim -\frac{1}{B_{a}}%
\sum_{l=0}^{\infty }\frac{(-1)^{l}w_{l}(\nu )}{u^{l+1}},  \label{Knuratas}
\end{equation}%
with the same coefficients $w_{l}(\nu )$ as in Eq. (\ref{Inuratas}). Now Eq.
(\ref{FRs}) is written in the form%
\begin{equation}
\begin{split}
F_{a}^{{\rm (R)}}(s) = & -\frac{(4\pi
)^{\frac{1-D}{2}}k_{D}^{D-1}(\mu
z_{a})^{-1-s}}{\Gamma \left( -\frac{s}{2}\right) \Gamma \left( \frac{D+1+s}{2%
}\right) }\left\{ B_{a}\int_{0}^{1}du\,u^{D+s}\frac{K_{\nu }(u)}{\bar{K}%
_{\nu }^{(a)}(u)}\right.  \\
&+\left. \int_{1}^{\infty }du\,u^{D+s}\left[ B_{a}\frac{K_{\nu }(u)}{\bar{K}%
_{\nu }^{(a)}(u)}+\sum_{l=0}^{N}\frac{(-1)^{l}w_{l}(\nu )}{u^{l+1}}\right]
+\sum_{l=0}^{N}\frac{(-1)^{l}w_{l}(\nu )}{D+s-l}\right\} .  \label{FRs1}
\end{split}
\end{equation}%
For $N\geq D-1$ the integrals on the right of this formula are
finite at the point $s=-1$ and $ F_{a}^{{\rm (R)}}(s)$ has a
simple pole at $s=-1$ presented by the summand $l=D-1$ in the last
sum. Hence, the expression for the field square contains pole and
finite parts:
\begin{equation}
\langle 0|\varphi ^{2}|0\rangle _{z=z_{a}}^{{\rm (R)}}=\langle \varphi
^{2}\rangle _{z=z_{a},{\rm p}}^{{\rm (R)}}+\langle \varphi ^{2}\rangle
_{z=z_{a},{\rm f}}^{{\rm (R)}},  \label{phi2Rpf}
\end{equation}%
where the separate contributions are obtained by Laurent-expanding
expression (\ref{FRs1}) near the point $s=-1$:
\begin{equation}
\langle \varphi ^{2}\rangle _{z=z_{a},{\rm p}}^{{\rm (R)}} =
2(-1)^{D}k_{D}^{D-1}w_{D-1}(\nu )\frac{\beta _{D+1}}{s+1},
\label{phi2Rpf1}
\end{equation}
and
\begin{equation}
\begin{split}
\langle \varphi ^{2}\rangle _{z=z_{a},{\rm f}}^{{\rm (R)}} = & -2
k_{D}^{D-1}\beta _{D+1} \left\{
B_{a}\int_{0}^{1}du\,u^{D-1}\frac{K_{\nu }(u)}{\bar{K}_{\nu
}^{(a)}(u)}\right.  \\
&+\int_{1}^{\infty }du\,u^{D-1}\left[ B_{a}\frac{K_{\nu
}(u)}{\bar{K}_{\nu
}^{(a)}(u)}+\sum_{l=0}^{N}\frac{(-1)^{l}w_{l}(\nu
)}{u^{l+1}}\right]
+\sum_{\substack{l=0\\ l\neq D-1}}^{N}\frac{(-1)^{l}w_{l}(\nu )}{D-l-1}  \\
&+\left. (-1)^{D}w_{D-1}(\nu )\left[ \ln (\mu
z_{a})+\frac{1}{2}\psi \left( \frac{D}{2}\right) -\frac{1}{2}\psi
\left( \frac{1}{2}\right) \right] \right\} .  \label{phi2Rpf2}
\end{split}
\end{equation}%
With the help of these formulae we obtain the similar
decomposition for the surface energy density in the R-region:
\begin{eqnarray}
&& \varepsilon _{a}^{{\rm (surf)(R)}}=\varepsilon _{a,{\rm
p}}^{{\rm (surf)(R)}}+\varepsilon _{a,{\rm f}}^{{\rm (surf)(R)}},
\label{epspfR} \\
&& \varepsilon _{a,{\rm s}}^{{\rm (surf)(R)}}=-\left[ \zeta
k_{D}-(2\zeta -1/2)\tilde{A}_{a}/\tilde{B}_{a}\right] \langle
\varphi ^{2}\rangle _{z=z_{a},{\rm s}}^{{\rm (R)}},
\end{eqnarray}
with ${\rm s=p,f}$. In the minimal subtraction scheme the pole
term appearing in Eq. (\ref{epspfR}) is omitted. This corresponds
to the principal part prescription of Ref. \cite{Blau88} for the
total Casimir energy. Note that the structure of the pole terms in
Eqs. (\ref{epspf}) and (\ref{epspfR}) allows us to absorb them
into the corresponding counterterms. There is the freedom to
perform finite renormalizations since the renormalization
counterterms shall be fixed by imposing some renormalization
conditions (for the detailed  discussion of the renormalization
procedure for the vacuum energy in the Randall-Sundrum braneworld
see Refs. \cite{Gold00,Garr01,Flac01a}).

\begin{figure}[tbph]
\begin{center}
\begin{tabular}{cc}
\epsfig{figure=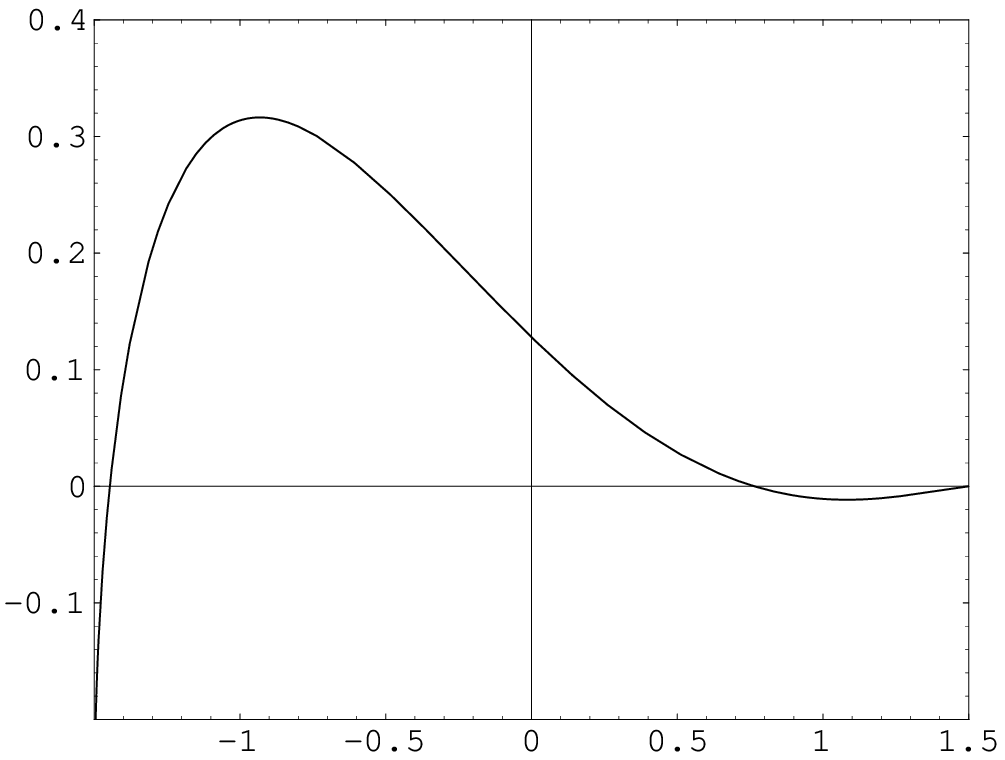,width=6.5cm,height=5cm}& \quad
\epsfig{figure=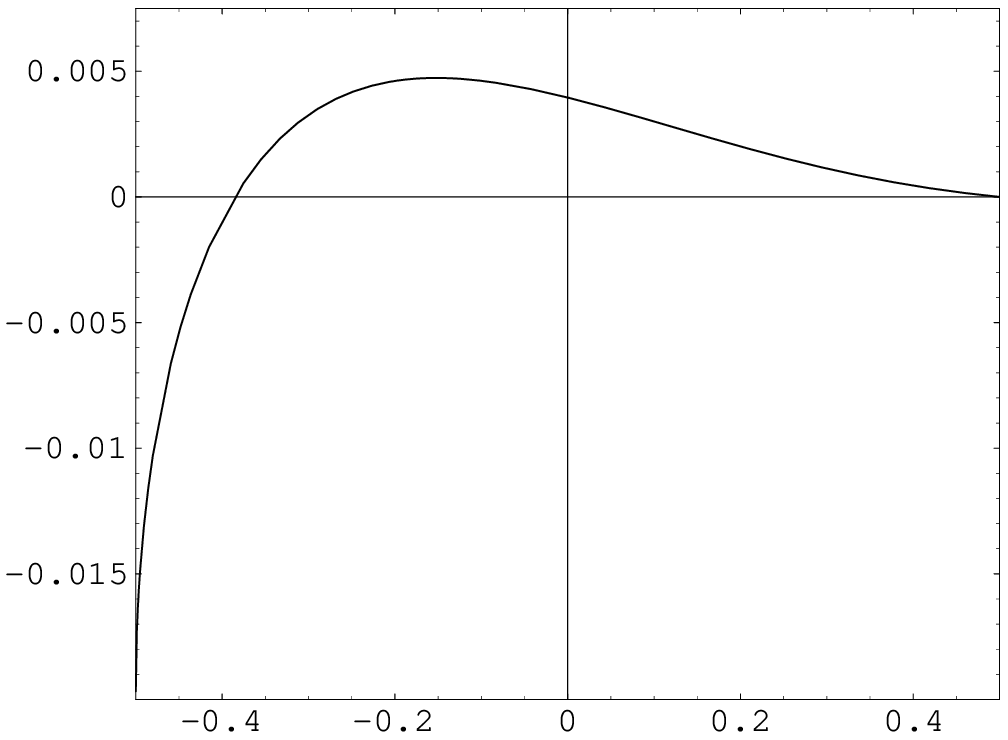,width=6.5cm,height=5cm}
\end{tabular}
\end{center}
\caption{Surface energy density, $k_{D}^{-D}\varepsilon
_{a}^{{\mathrm{(surf)}}}$ [see formula (\ref{epsLR1})], induced on
a single plate as a function on the ratio $A_a/B_a$ for massless
minimally (left panel) and conformally (right panel) coupled
scalar fields in $D=3$.} \label{fig1}
\end{figure}

The total surface energy density for a single brane at $y=a$ is
obtained by summing the
contributions from the L- and R-regions:%
\begin{equation}
\varepsilon ^{{\mathrm{(surf)(LR)}}}_a=\varepsilon _{a}^{{\rm
(surf)(L)}}+\varepsilon _{a}^{{\rm (surf)(R)}}. \label{epsLR}
\end{equation}%
Now comparing the pole parts for the separate regions, we see that in odd
spatial dimensions these parts cancel out. In particular, taking $N=D-1$,
for the total surface energy density one obtains the formula%
\begin{equation}
\begin{split}
\varepsilon _{a}^{{\rm (surf)(LR)}}= & k_{D}^{D}\beta _{D+1}\left[
2\zeta B_a-(4\zeta -1)\tilde{A}_{a}\right]\left\{
\int_{0}^{1}du\,u^{D-1}\left[ \frac{I_{\nu }(u)}{\bar{I}_{\nu }^{(a)}(u)%
}+\frac{K_{\nu }(u)}{\bar{K}_{\nu }^{(a)}(u)}\right] \right.  \label{epsLR1}
\\
&+ \int_{1}^{\infty }du\,u^{D-1}\left[ \frac{I_{\nu }(u)}{\bar{I}%
_{\nu }^{(a)}(u)}+\frac{K_{\nu }(u)}{\bar{K}_{\nu }^{(a)}(u)}%
-\frac{2}{B_{a}}\sum_{l=0}^{\frac{D-3}{2}}\frac{w_{2l+1}(\nu
)}{u^{2l+2}}\right] \\
& \left.
-\frac{2}{B_{a}}\sum_{l=0}^{\frac{D-3}{2}}\frac{w_{2l+1}(\nu
)}{D-2l-2}\right\} ,
\end{split}
\end{equation}%
where the coefficients $w_{l}(\nu )$ are defined by relation (\ref%
{Inuratas}). Note that this quantity does not depend on the
renormalization scale $\mu $ and the position of the brane. In
Fig. \ref{fig1} we have plotted the surface energy density
(\ref{epsLR1}) as a function on the ratio $A_a/B_a$ for massless
minimally (left panel) and conformally (right panel) coupled
scalar fields in $D=3$.

\section{Surface densities for two-branes geometry and induced
cosmological constant} \label{sec:2brane}

In this section we investigate the surface densities generated on
a brane by the presence of the second brane. These quantities are
finite for all nonzero values of the interbrane separation and are
not affected by finite renormalizations. Below we will concentrate
on the energy densities as the corresponding stresses on the
branes are directly obtained from the equation of state
(\ref{eqstate}). On the base of relation (\ref{Fab}), for the
region $z_a\leq z\leq z_b$ the vacuum expectation value of the
field square on the brane at $y=j$ ($j=a,b$) can be presented in
the form
\begin{equation}
\langle 0|\varphi ^{2}(x)|0\rangle _{z=z_{j}}=\langle 0|\varphi
^{2}(x)|0\rangle _{z=z_{j}}^{{\rm
(J)}}-2k_{D}^{D-1}z_{j}^{D}B_{j}^{2}\beta _{D+1}\int_{0}^{\infty
}du\,u^{D-1}\Omega _{j\nu }(uz_{a},uz_{b}),  \label{phi22pln}
\end{equation}%
where $\langle 0|\varphi ^{2}(x)|0\rangle _{z=z_{j}}^{{\rm (J)}}$
is the corresponding quantity for a single brane at $y=j$ when the
second brane is absent, ${\rm J=R}$ \ for $j=a$ and ${\rm J=L}$ \
for $j=b$. The second term on the right of Eq. (\ref{phi22pln})
may also be obtained by using the expression for the corresponding
Wightman function derived in Ref. \cite{Saha04a} in the
coincidence limit and evaluating its value on the branes. With the
help of formulae (\ref{tauj}), (\ref{phi22pln}), the surface
energy density on the brane at $z=z_j$ is presented as the sum
\begin{equation}
\varepsilon ^{{\mathrm{(surf)}}}_j=\varepsilon
^{{\mathrm{(surf)(J)}}}_j +\Delta \varepsilon
^{{\mathrm{(surf)}}}_j, \label{emt2pl2}
\end{equation}%
where $\varepsilon ^{{\mathrm{(surf)(J)}}}_j$ is the surface
energy density induced on the corresponding surface of a single
brane at $y=j$ when the second brane is absent (see previous
section), and the term
\begin{equation}
\Delta \varepsilon ^{{\mathrm{(surf)}}}_j=\left[ 2\zeta
B_{j}-\left( 4\zeta -1\right) \tilde{A}_{j}\right]
n^{(j)}(k_{D}z_{j})^{D}B_{j}\beta _{D+1}\int_{0}^{\infty
}du\,u^{D-1}\Omega _{j\nu }(uz_{a},uz_{b})  \label{emt2pl3}
\end{equation}
is the energy density induced by the presence of the second brane.
Note that this part is located on the surface $z=z_{a}+0$ for the
brane at $z=z_a$ and on the surface $z=z_b-0$ for the brane at
$z=z_b$. On the surfaces $z=z_a-0$ and $z=z_b+0$ the surface
densities are the same as for single branes. The expression on the
right of Eq. (\ref{emt2pl3}) is finite for all values
$z_{a}<z_{b}$ and is a function on the ratio $z_b/z_a$ only. Note
that this ratio is related to the interbrane distance by the
formula
\begin{equation}\label{zbza}
z_b/z_a=e^{k_D(b-a)}.
\end{equation}
The main contribution into the integral in Eq. (\ref{emt2pl3})
comes from the modes with $u\lesssim (z_b-z_a)^{-1}$ and the
contribution of higher modes is exponentially suppressed. Due to
this suppression, the same results for the induced densities will
be obrained in the model where instead of externally imposed
boundary condition the fluctuating field is coupled  to a smooth
background potential that implements the boundary condition in a
certain limit \cite{Grah03}.

Using the Wronskian for the Bessel modified functions, it can be
seen that
\begin{equation} \label{Omtrans}
\left[ B_j^2(x^2z_j^2+\nu ^2)-A_j^2\right] \Omega _{j\nu } (x
z_a,x z_b)=n^{(j)}z_j\frac{\partial }{\partial z_j}\ln \left|
1-\frac{\bar I_{\nu }^{(a)}(xz_a)\bar K_{\nu }^{(b)}(xz_b)}{\bar
I_{\nu }^{(b)}(xz_b)\bar K_{\nu }^{(a)}(xz_a)}\right| ,
\end{equation}
for $j=a,b$. This allows us to write the expression
(\ref{emt2pl3}) for the surface energy density in another
equivalent form:
\begin{equation}
\begin{split}
\Delta \varepsilon ^{{\mathrm{(surf)}}}_j = & k_{D}^D
z_{j}^{D+1}B_{j}\beta _{D+1} \int _{0}^{\infty }du\, u^{D-1}
\frac{2\zeta B_j+(1-4\zeta )\tilde A_j}{B_j^2(u^2z_j^2+
\nu ^2)-A_j^2}\\
&\times  \frac{\partial }{\partial z_j}\ln \left| 1-\frac{\bar
I_{\nu }^{(a)}(uz_a)\bar K_{\nu }^{(b)}(uz_b)}{\bar I_{\nu
}^{(b)}(uz_b)\bar K_{\nu }^{(a)}(uz_a)}\right|  . \label{emt2pl4}
\end{split}
\end{equation}
This form of the surface energy density will be used below in the
discussion of the energy balance.

Let us consider the limiting cases of the part (\ref {emt2pl3}).
For large values of AdS radius to compared with the interbrane
distance, $ k_{D}(b-a)\ll 1$, the main contribution to the
integral on the right of Eq. (\ref{emt2pl3}) comes from the large
values of $uz_a\sim [k_D(b-a)]^{-1}$. Assuming that $\tilde
B_a/(b-a)$ and $m(b-a)$ are fixed we see that the order of the
Bessel modified functions is large. Replacing these functions by
their uniform asymptotic expansions for large values of the order
(see \cite{abramowiz}), one obtains
\begin{equation}\label{EpssmallkD}
\Delta \varepsilon ^{{\mathrm{(surf)}}}_j \approx
2n^{(j)}(1-4\zeta ) \tilde A_j \tilde B_j\beta
_{D+1}\int_{m}^{\infty }du
\frac{u^2(u^2-m^2)^{\frac{D}{2}-1}(\tilde A_j^2-\tilde
B_j^2u^2)^{-1}}{\frac{(\tilde A_a-\tilde B_a u)(\tilde A_b+\tilde
B_b u)}{(\tilde A_a+\tilde B_a u)(\tilde A_b-\tilde B_b u)}
e^{2u(b-a)}-1} .
\end{equation}
The expression on the right is the corresponding surface energy
for two parallel Robin plates in the Minkowski spacetime. It can
be easily checked that for a massless scalar field the sum $\Delta
\varepsilon ^{{\mathrm{(surf)}}}_a + \Delta \varepsilon
^{{\mathrm{(surf)}}}_b$ evaluated from Eq. (\ref{EpssmallkD})
coincides with the result obtained in Ref. \cite{Rome02}. In the
limit of small interbrane separation, $(b-a)\to 0$, for a fixed
$k_D$, by using the asymptotic formulae for the Bessel modified
functions for large values of the argument, from the general
formula in $D>2$ dimensions we find
\begin{equation}\label{Epssmalldist}
\Delta \varepsilon ^{{\mathrm{(surf)}}}_j \approx - \left[ 2\zeta
-(4\zeta -1)\tilde A_j/B_j\right] \frac{n^{(j)}\sigma _j k_D\zeta
_R(D-1)\Gamma \left( \frac{D+1}{2}\right) }{2^{D-1}\pi
^{\frac{D+1}{2}}(D-1)(b-a)^{D-1}} ,
\end{equation}
where $\zeta _R(x)$ is the Riemann zeta function, $\sigma _{j}=1$
for $|B_a/A_a|, |B_b/A_b|\gg k_D(b-a)$, and $\sigma _j=2^{2-D}-1$
for $|B_j/A_j|\gg k_D(b-a)$ and $B_l/A_l=0$, with $l=b$ for $j=a$
and $l=a$ for $j=b$. We see that for small interbrane distances
the sign of the induced surface energy density is determined by
the factor $\left[ 2\zeta -(4\zeta -1)\tilde A_j/B_j\right] $ and
this sign is different for two cases of $\sigma _j$.

Now we turn to the asymptotic behavior of the surface energy
density (\ref{emt2pl3}) for large distances between the branes, $
k_{D}(b-a)\gg 1$. Introducing a new integration variable
$x=uz_{b}$, by making use of formulae for the Bessel modified
functions for small values of the argument, and assuming
$|A_{a}|\neq |B_{a}|\nu $, to the leading order we receive
\begin{equation}
\Delta \varepsilon ^{{\mathrm{(surf)}}}_a \approx \left( \frac{z_{a}}{2z_{b}%
}\right) ^{D+2\nu }\frac{4k_{D}^{D}B_{a}\left[ 2\zeta B_{a}-\left(
4\zeta
-1\right) \tilde{A}_{a}\right] }{\pi ^{\frac{D}{2}}\Gamma \left( \frac{D}{2}%
\right) \Gamma ^{2}(\nu )\left( A_{a}-B_{a}\nu \right) ^{2}}\int_{0}^{\infty
}dx\,x^{D+2\nu -1}\frac{\bar{K}_{\nu }^{(b)}(x)}{\bar{I}_{\nu }^{(b)}(x)},
\label{Deltatau3}
\end{equation}%
and%
\begin{equation}
\Delta \varepsilon ^{{\mathrm{(surf)}}}_b =k_{D}^{D}\left( \frac{z_{a}}{%
z_{b}}\right) ^{2\nu }\frac{B_{a}\nu +A_{a}}{B_{a}\nu
-A_{a}}f_{\nu }^{(b)}, \label{Deltatau4}
\end{equation}
where we have introduced the notation
\begin{equation}\label{fnub}
f_{\nu }^{(b)}= \frac{B_{b}\left[ 2\zeta B_{b}-\left( 4\zeta
-1\right) \tilde{A}_{b} \right] }{2^{2\nu +D-1}\pi
^{\frac{D}{2}}\Gamma \left( \frac{D}{2}\right) \nu \Gamma ^{2}(\nu
)}\int_{0}^{\infty }dx\,\frac{x^{D+2\nu -1}}{\bar{I}_{\nu
}^{(b)2}(x)} .
\end{equation}
For  $|A_{a}|= |B_{a}|\nu $ it is necessary to take into account
the next terms in the corresponding expansions of the Bessel
modified functions for small values of the argument. The integral
in Eq. (\ref{Deltatau3}) is negative for small values of the ratio
$A_b/B_b$ and is positive for large values of this ratio. As it
follows from Eq. (\ref{Deltatau4}), the sign of the quantity
$\Delta \varepsilon ^{{\mathrm{(surf)}}}_b $ for large interbtane
separations is determined by the combination $(B_a^2\nu
^2-A_a^2)[2\zeta +(1-4\zeta )\tilde A_b/B_b]$ of the coefficients
in the boundary conditions.

\begin{figure}[tbph]
\begin{center}
\begin{tabular}{cc}
\epsfig{figure=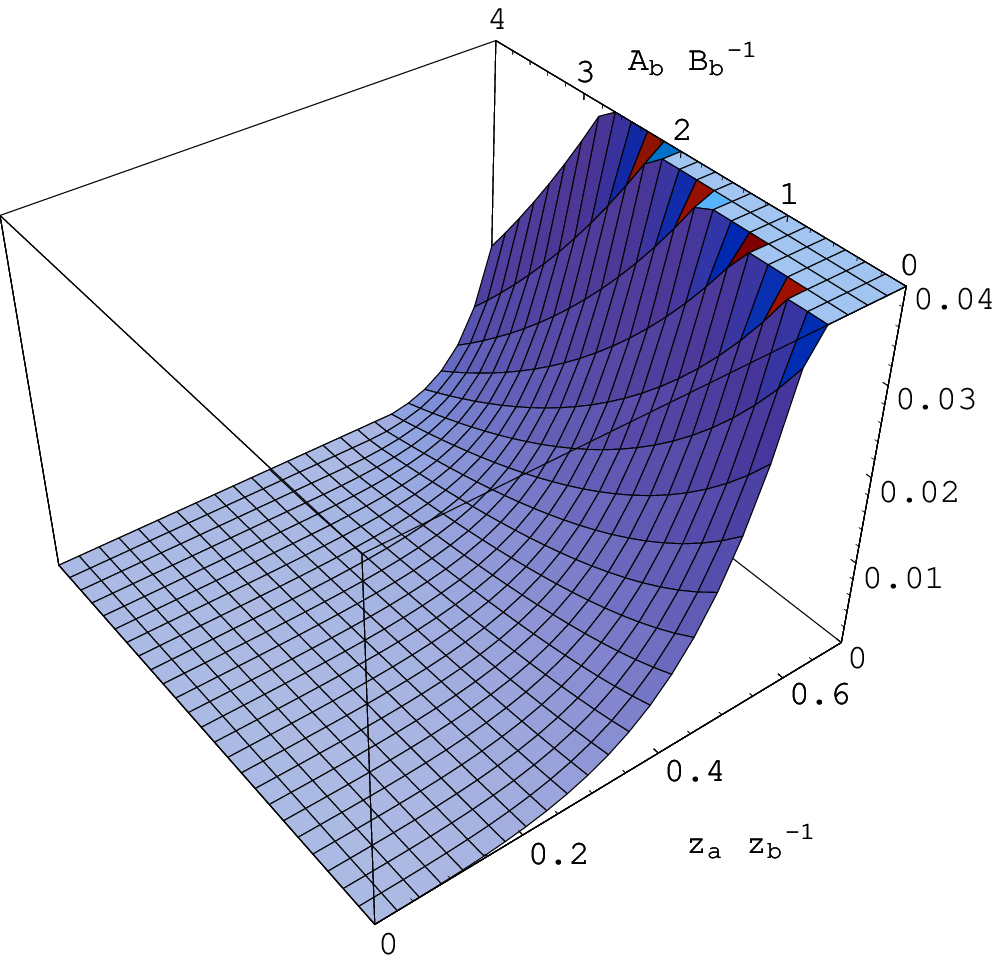,width=6.5cm,height=6cm}& \quad
\epsfig{figure=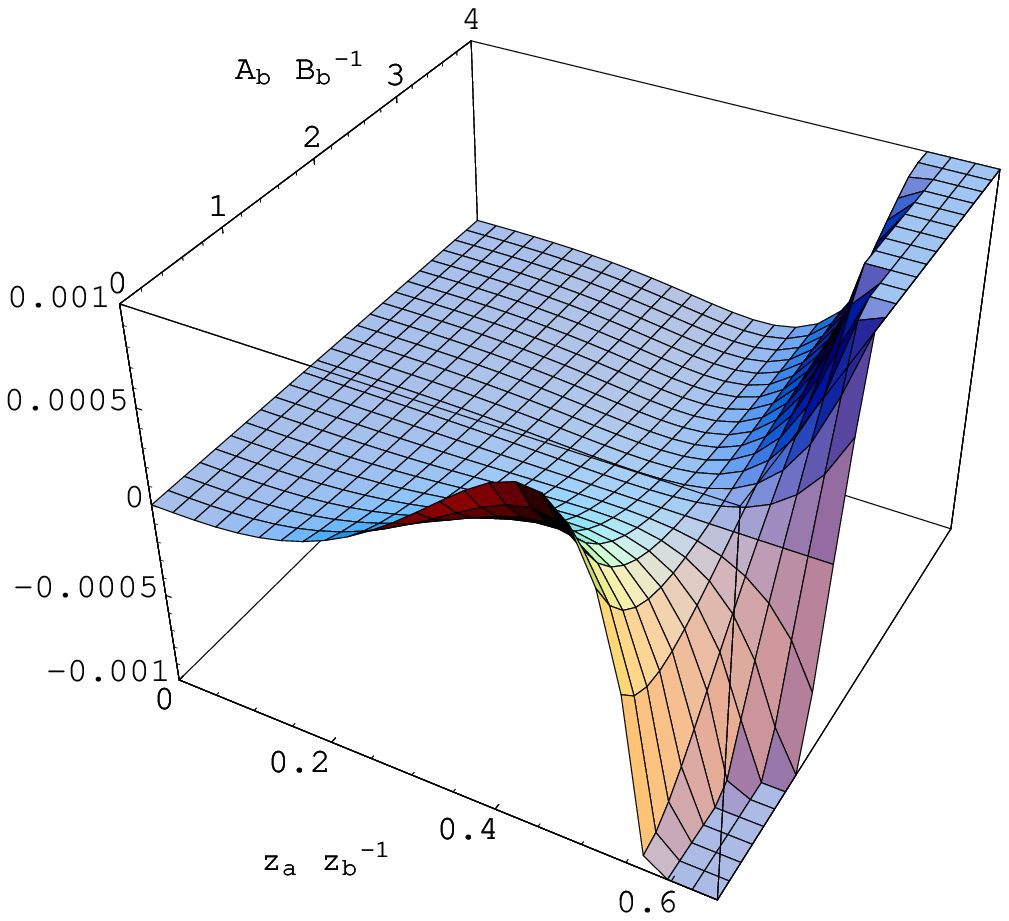,width=6.5cm,height=6cm}
\end{tabular}
\end{center}
\caption{The surface energy densities $(z_b/z_j)^{D}\Delta
\varepsilon ^{(j)}/k_D^{D}$, $j=a,b$, as functions on $z_a/z_b$
and $A_b/B_b$ for a massless minimally coupled scalar field with
$A_a/B_a=-2$ in $D=3$ induced on the branes at $z=z_a$ (left
panel) and $z=z_b$ (right panel).} \label{fig2}
\end{figure}

For large values of the mass with $m\gg k_D$ and $m(b-a)\gg 1$, as
we see from Eq. (\ref{nu}), one has $\nu \sim m/k_D\gg 1$.
Introducing a new integration variable $u=\nu y$ and using the
uniform asymptotic expansions for the Bessel modified functions,
to the leading order from (\ref{emt2pl3}) one receives
\begin{equation}\label{largem}
\Delta \varepsilon ^{{\mathrm{(surf)}}}_j \approx n^{(j)}
\frac{2\zeta B_j-(4\zeta -1)\tilde A_j}{\left[ 2\pi
(z_b^2/z_a^2-1)\right] ^{D/2}} \frac{(A_l+n^{(l)}m\tilde B_l)(m
k_D)^{D/2}m \tilde B_j}{(A_l-n^{(l)}m\tilde
B_l)(A_j-n^{(j)}m\tilde B_j)^2}e^{-2m(b-a)} ,
\end{equation}
where $l=a$ for $j=b$ and $l=b$ for $j=a$, and we keep $m\tilde
B_l$ an arbitrary. As we could expect in this case the induced
surface densities are exponentially suppressed. In Fig. \ref{fig2}
we have plotted the surface energy densities determined by Eq.
(\ref{emt2pl3}) as functions on $z_a/z_b$ and $A_b/B_b$ for
$A_a/B_a=-2$.

To discuss the physics from the point of view of a $D$-dimensional
observer residing on the brane $y=j$, it is convenient to
introduce rescaled coordinates $x_{j}'{}^{\mu }$ in accordance
with $x_{j}'{}^{\mu }=e^{-k_Dj}x_{j}^{\mu }$. With these
coordinates the warp factor in the metric is equal to 1 at the
brane $y=j$ and they are physical coordinates for an observer on
this brane. Now after the dimensional reduction of the action, by
the way similar to that in the Randall-Sundrum braneworld (see,
e.g., \cite{Ruba01}), it can be seen that $D$-dimensional Newton's
constant $G_{Dj}$ measured by an observer on the brane at $y=j$ is
related to the fundamental $(D+1)$-dimensional Newton's constant
$G_{D+1}$ by the formula (for the gravitational equation on the
brane see, e.g., \cite{Maar03},\cite{Shir00})
\begin{equation}\label{GDj}
G_{Dj}=\frac{(D-2)k_DG_{D+1}}{e^{(D-2)k_D(b-a)}-1}
e^{(D-2)k_D(b-j)}.
\end{equation}
Note that in the orbifolded version of the model an additional
factor 2 appears in the denominator of the expression on the
right. For large interbrane distances one has $G_{Da}\sim
k_DG_{D+1}$, $G_{Db}\sim k_DG_{D+1}e^{(2-D)k_D(b-a)}$, and the
gravitational interactions on the brane $y=b$ are exponentially
suppressed. This feature is used in the Randall-Sundrum model to
address the hierarchy problem. Now we will show that this
mechanism also allows to obtain a naturally small cosmological
constant generated by vacuum quantum fluctuations.

The surface energy density (\ref{emt2pl3}) corresponds to the
gravitational source of the cosmological constant type induced on
the brane at $z=z_j$ by the presence of the second brane. Within
two brane geometry discussed in this section, from the point of
view of an observer living on the brane at $z=z_j$ the effective
cosmological constant induced by the second brane is determined by
the relation
\begin{equation}\label{effCC}
\Lambda _{Dj}=8\pi G_{Dj} \Delta \varepsilon
^{{\mathrm{(surf)}}}_j =\frac{8\pi \Delta \varepsilon
^{{\mathrm{(surf)}}}_j }{M^{D-2}_{Dj}},
\end{equation}
where $M_{Dj}$ is the $D$-dimensional effective Planck mass scale
for an observer on the brane. By using relation (\ref{GDj}), for
the numerical estimates it is convenient to present the ratio of
the induced cosmological constant (\ref{effCC}) to the
corresponding Planck scale quantity in the brane universe in the
form
\begin{equation}\label{LambDj}
\frac{\Lambda _{Dj}}{8\pi G_{Dj}M_{Dj}^D}=\frac{\Delta \varepsilon
^{{\mathrm{(surf)}}}_j}{k_D^D}\left( \frac{z_b}{z_j}\right) ^D
\left( \frac{k_D}{M_{D+1}}\right) ^{\frac{D(D-1)}{D-2}} \left[
\frac{D-2}{(z_b/z_a)^{D-2}-1}\right] ^{\frac{D}{D-2}} ,
\end{equation}
where $M_{D+1}$ is the fundamental $(D+1)$-dimensional Planck
mass, $G_{D+1}=M_{D+1}^{1-D}$. For large interbrane distances, by
taking into account formulae (\ref{Deltatau3}), (\ref{Deltatau4}),
one obtains the following estimate
\begin{equation}\label{LambDj1}
\frac{\Lambda _{Dj}}{8\pi G_{Dj}M_{Dj}^D}\sim \left(
\frac{z_a}{z_b}\right) ^{D+2\nu } \left(
\frac{k_D}{M_{D+1}}\right) ^{\frac{D(D-1)}{D-2}} ,
\end{equation}
showing that this ratio is of the same order of magnitude for both
branes. However, due to the large hierarchy between the Newton's
constants on the branes, for the ratio of the induced cosmological
constants one has $\Lambda _{Db}/\Lambda _{Da}\sim (z_b/z_a)^2$.

\begin{figure}[tbph]
\begin{center}
\begin{tabular}{cc}
\epsfig{figure=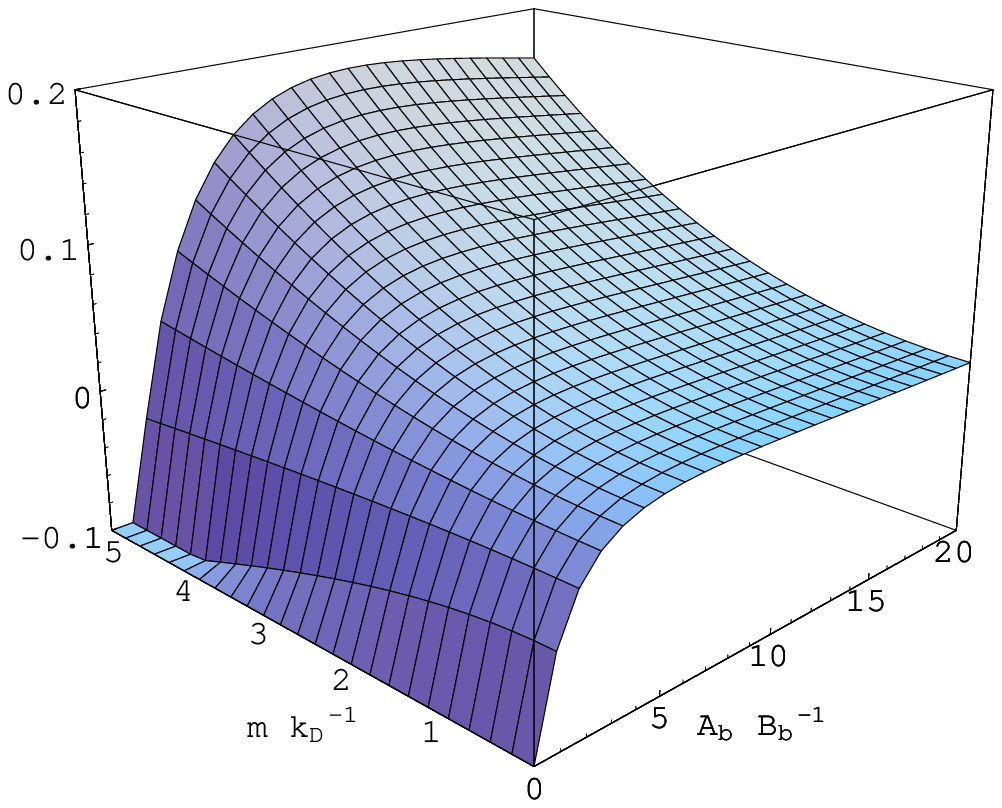,width=6.5cm,height=6cm}& \quad
\epsfig{figure=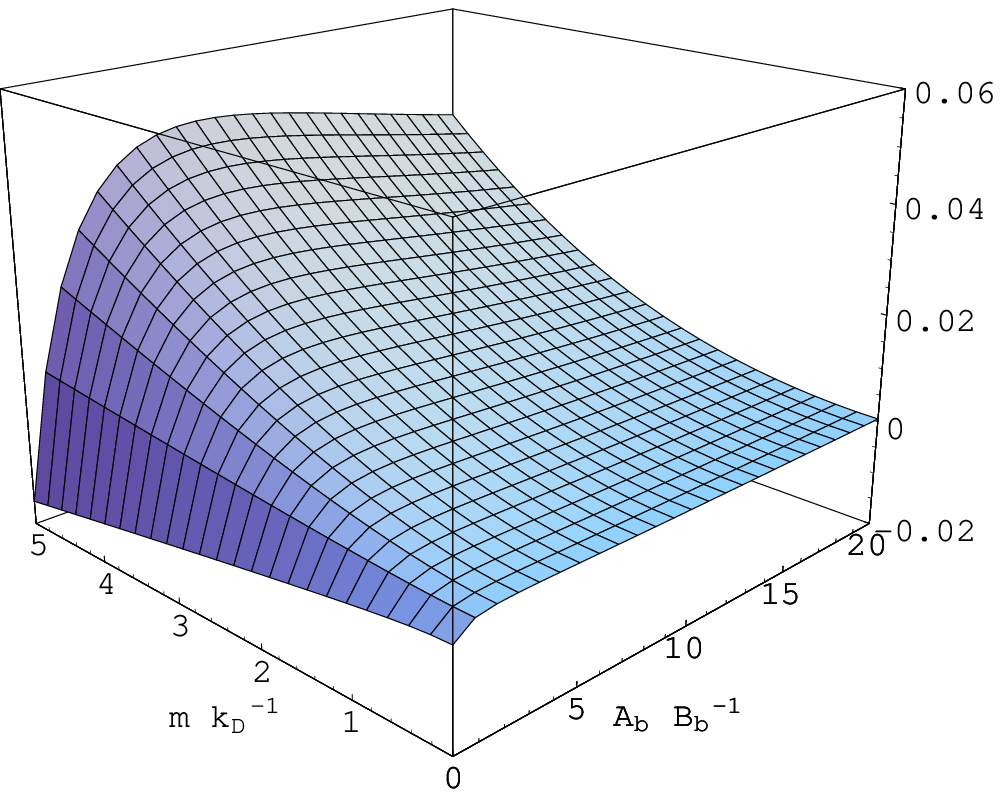,width=6.5cm,height=6cm}
\end{tabular}
\end{center}
\caption{The quantity $f_{\nu }^{(b)}$ defined by Eq. (\ref{fnub})
as a function on the ratio $A_b/B_b$ and $m/k_D$ in the model with
$D=4$ for minimally (left panel) and conformally (right panel)
coupled scalar fields.} \label{fig3}
\end{figure}

In the Randall-Sundrum $(D+1)$-dimensional braneworld the brane at
$z=z_b$ corresponds to the visible brane. For large interbrane
distances by taking into account Eq. (\ref{Deltatau4}), for the
ratio of the induced cosmological constant (\ref{effCC}) to the
corresponding Planck scale quantity in the corresponding brane
universe one obtains
\begin{equation}\label{LambDb}
\frac{\Lambda _{Db}}{8\pi G_{Db}M_{Db}^D}\approx \left(
\frac{z_a}{z_b}\right) ^{D+2\nu } \left(
\frac{k_D}{M_{D+1}}\right) ^{\frac{D(D-1)}{D-2}} \frac{B_{a}\nu
+A_{a}}{B_{a}\nu -A_{a}}f_{\nu }^{(b)},
\end{equation}
where the function $f_{\nu }^{(b)}$ is defined by Eq.
(\ref{fnub}). As this function determines the value of the induced
cosmological constant for large interbrane distances, in Figure
\ref{fig3} we have plotted its dependence on the ratio of the
coefficients in the boundary condition on the visible brane and on
the mass of the scalar field for minimally and conformally coupled
cases.

The higher dimensional Planck mass $M_{D+1}$ and AdS inverse
radius $k_D$ are two fundamental energy scales in the
$(D+1)$-dimensional AdS bulk space which in the Randall-Sundrum
model are usually assumed to be of the same order, $k_D\sim
M_{D+1}$ (see, e.g., \cite{Ruba01}). In this case from Eq.
(\ref{LambDb}) one obtains the cosmological constant which is
exponentially suppressed to compared with the corresponding Planck
scale quantity on the visible brane. In the original
Randall-Sundrum braneworld with $D=4$, $M_{D+1}\sim TeV$,
$M_{Db}=M_{{\mathrm{Pl}}}\sim 10^{16}\, TeV $, to account for the
observed hierarchy between the gravitational and electroweak
scales, in accordance with Eq. (\ref{GDj}) we need $z_b/z_a\sim
10^{16}$ (by formula (\ref{zbza}) this corresponds to the
interbrane distance about 37 times larger than the AdS radius
$k_D^{-1}$). In the case of a scalar field with the mass
$|m^2|\lesssim k_D^2$ (recall that on AdS bulk the parameter $m^2$
can also be negative), for the induced cosmological constant this
leads to the estimate which is of the right order of magnitude
with the value implied by the cosmological observations.

\section{Energy balance}
\label{sec:enbal}

In this section we will consider the balance between the separate
parts of the vacuum energy. In the region between the branes the
total vacuum energy per unit coordinate volume on the branes is
the sum of zero-point energies of elementary oscillators:
\begin{equation}\label{toten1}
    E=\frac{1}{2}\int \frac{d^{D-1}{\mathbf{k}}}{(2\pi
    )^{D-1}}\sum _{n=1}^{\infty }(k^2+u_{\nu ,n})^{1/2},
\end{equation}
where $u_{\nu ,n}$ are roots of equation (\ref{cnu}). The total
vacuum energy in the bulk region $z_a\leq z\leq z_b$ is obtained
by the integration of ${^{0}_{0}}$-component of the volume
energy-momentum tensor over this region:
\begin{equation}\label{Evol}
E^{{\mathrm{(vol)}}}=\int d^{D+1}x\, \sqrt{|g|}\langle
0|T^{{\mathrm{(vol)}}0}_0|0\rangle .
\end{equation}
Using the formula for the volume energy-momentum tensor from Ref.
\cite{Saha04a}, we can see that this energy differs from
(\ref{toten1}) (see Ref. \cite{Saha03} for the discussion in
general case of bulk and boundary geometries). This difference is
due to the presence of the surface energy located on the branes.
The surface energy per unit coordinate volume on the brane,
$E^{{\mathrm{(surf)}}}$, is related to the surface densities from
Eq. (\ref{tauj1}) by the formula
\begin{equation}\label{Esurfcoord}
E^{{\mathrm{(surf)}}}=\sum_{j=a,b}\frac{\varepsilon
^{{\mathrm{(surf)}}}_j}{(k_Dz_j)^{D}}.
\end{equation}
Now it can be easily checked that the total vacuum energy is the
sum of the volume part and the surface part:
\begin{equation}\label{EvolEsurf}
E=E^{{\mathrm{(vol)}}}+E^{{\mathrm{(surf)}}}.
\end{equation}

Expression (\ref{toten1}) may be evaluated using the procedure
which is basically the same as that used in Refs. \cite{Eliz93}
(see \cite{Kirs01,Byts03} for reviews) for studying the Casimir
effects in the geometry of spherical boundaries. Within the
framework of the Randall-Sundrum braneworld, this has been done in
Refs. \cite{Toms00,Gold00,Flac01b} by the dimensional
regularization method and in Ref. \cite{Garr01} by the zeta
function technique. Refs. \cite{Toms00,Gold00,Garr01} consider the
case of a minimally coupled scalar field in $D=4$, and the case of
arbitrary $\zeta $ and $D$ with zero mass terms $c_a$ and $c_b$ is
calculated in Ref. \cite{Flac01b}. Here we briefly outline the
zeta function approach in the general case. Instead of
(\ref{toten1}) we introduce the following zeta function
\begin{equation}\label{zetatot1}
    \zeta (s)=\mu ^{s+1}\int \frac{d^{D-1}{\mathbf{k}}}{(2\pi
    )^{D-1}}\sum _{n=1}^{\infty }(k^2+u_{\nu ,n})^{-s/2},
\end{equation}
where, as above, the parameter $\mu $ with dimension of mass is
introduced by dimensional reasons. Evaluating the integral over
${\mathbf{k}}$ one receives
\begin{equation}\label{zetatot2}
    \zeta (s)=\frac{\mu ^{s+1}}{(4\pi )^{\frac{D-1}{2}}}
    \frac{\Gamma \left( \frac{s-D+1}{2}\right)}{\Gamma \left( \frac{s}{2}\right) }
    \sum _{n=1}^{\infty }u_{\nu ,n}^{D-s-1}.
\end{equation}
We need to perform the analytic continuation of the sum on the
right of this formula to the neighborhood of $s=-1$. Transforming
it into a contour integral and by deforming the contour
appropriately, the total energy in the region $z_a\leq z\leq z_b$
is presented in the form
\begin{equation}\label{totendec}
E=\frac{1}{2}\zeta (s)|_{s=-1}=E^{(a)}_{z\geq z_a}+E^{(b)}_{z\leq
z_b}+\Delta E,
\end{equation}
where $E^{(a)}_{z\geq z_a}$ is the vacuum energy in the region
$z\geq z_a$ for a single brane at $z=z_a$, and $E^{(b)}_{z\leq
z_b}$ is the vacuum energy in the region $z\leq z_b$ for a single
brane at $z=z_b$. The interference term $\Delta E$ in this formula
is finite for all nonzero values of the interbrane separation (see
below, Eq. (\ref{DeltaE})) and the analytic continuation is needed
for single brane parts only. Under the assumption $B_b\neq 0$, for
the region $z\leq z_b$ the latter is given by the expression
\begin{equation}\label{Eb1}
E^{(b)}_{z\leq z_b}=\frac{\mu ^{s+1}(4\pi )^{(1-D)/2}}{2\Gamma
\left( \frac{s}{2}\right) \Gamma \left( \frac{D+1-s}{2}\right) }
\int_{0}^{\infty }du \, u^{D-s-1}\left. \frac{d}{du} \ln \left[
\Sigma _{\nu }^{(I,b)}(uz_b) \right] \right| _{s=-1} ,
\end{equation}
where we have introduced the notation
\begin{equation}\label{SigIb}
\Sigma _{\nu }^{(I,b)}(uz_b)=\sqrt{\frac{2\pi
}{uz_b}}\frac{e^{-uz_b}}{B_b}\bar I_{\nu }^{(b)}(uz_b),
\end{equation}
and, as before, $|_{s=-1}$ is understood in the sense of the
analytic continuation. The expression for the energy
$E^{(a)}_{z\geq z_a}$ is obtained from Eq. (\ref{SigIb}) by the
replacement
\begin{equation}\label{SigKa}
\Sigma _{\nu }^{(I,b)}(uz_b)\to \Sigma _{\nu }^{(K,a)}(uz_a)=
-\sqrt{\frac{2}{\pi uz_a}}\frac{e^{uz_a}}{B_a}\bar K_{\nu
}^{(a)}(uz_a).
\end{equation}
After the analytic continuation we can see that $E^{(b)}_{z\leq
z_b}$ contains pole and finite parts,
\begin{equation}\label{Eb1pf}
E^{(b)}_{z\leq z_b}=E^{(b)}_{z\leq
z_b,{\mathrm{p}}}+E^{(b)}_{z\leq z_b,{\mathrm{f}}},
\end{equation}
with
\begin{equation}\label{Eb1p}
E^{(b)}_{z\leq z_b,{\mathrm{p}}} = -\frac{w_{D}^{(b)}(\nu ) \beta
_{D+1}}{z_b^{D}(s+1)},
\end{equation}
for the pole part, and
\begin{equation}
\begin{split}
E^{(b)}_{z\leq z_b,{\mathrm{f}}}  = & -\frac{\beta _{D+1}}{D
z_b^{D}} \left\{ \int_{0}^{1}du\, u^D \frac{d}{du}\ln \left(
\Sigma _{\nu }^{(I,b)}(u)\right)\right. \\
& +\int_{1}^{\infty }du\, u^D \frac{d}{du}\left[ \ln \left( \Sigma
_{\nu }^{(I,b)}(u)\right)
-\sum_{l=1}^{N}\frac{w_{l}^{(b)}(\nu )}{u^l}\right] \label{Eb1f} \\
& +\left. \sum_{\substack{l=1\\ l\neq
D}}^{N}\frac{lw_{l}^{(b)}(\nu )}{D-l}+Dw_{D}^{(b)}(\nu ) \left[
\ln (\mu z_b)+\frac{1}{2}\psi \left(
\frac{D}{2}+1\right)-\frac{1}{2}\psi \left( -
\frac{1}{2}\right)\right] \right\} ,
\end{split}
\end{equation}
for the finite part, with $N>D-1$. In these formulae,
$w_{l}^{(j)}(\nu )$ are the coefficients in the asymptotic
expansion of the function $\ln \left[ \Sigma _{\nu
}^{(I,j)}(u)\right] $ for large values of the argument:
\begin{equation}\label{Defwnul}
\ln \left[ \Sigma _{\nu }^{(I,j)}(u)\right] \sim
\sum_{l=1}^{\infty }\frac{w_{l}^{(j)}(\nu )}{u^l}, \quad j=a,b.
\end{equation}
These coefficients can be related to the coefficients in the
similar expansions for the Bessel modified functions (see, for
instance, \cite{abramowiz}). The corresponding formulae for the
energy $E^{(a)}_{z\geq z_a}$ are obtained from expressions
(\ref{Eb1p}) and (\ref{Eb1f}) by replacements (\ref{SigKa}) and
$w_{l}^{(b)}(\nu )\to (-1)^{l}w_{l}^{(a)}(\nu )$. The
renormalization of the divergences in the corresponding formulae
for the vacuum energies can be performed by using the brane
tensions (for a detailed discussion of the renormalization
procedure within the framework of the Randall-Sundrum model see,
e.g., Refs. \cite{Gold00},\cite{Garr01},\cite{Flac01a}).

As in the case of the surface energies, now we see that in the
calculation of the total vacuum energy for a single brane at
$z=z_a$ in odd spatial dimensions, including the contributions
from L- and R-regions, the pole parts of the energies cancel out
(assuming that the coefficients in the boundary conditions
(\ref{boundcond}) on the right and left surfaces are the same) and
we obtain a finite result. In particular, taking $N=D-1$, for this
energy one receives
\begin{equation}
\begin{split}\label{toten1br}
E^{(a)} = & -\frac{\beta _{D+1}}{Dz_a^D}\left\{ \int_{0}^{1}du\,
u^D \frac{d}{du}\ln \left(
\Sigma _{\nu }^{(I,a)}(u)\Sigma _{\nu }^{(K,a)}(u)\right)\right. \\
& +\int_{0}^{1}du\, u^D \frac{d}{du}\left[\ln \left( \Sigma _{\nu
}^{(I,a)}(u)\Sigma _{\nu }^{(K,a)}(u)\right)
-2\sum_{l=1}^{\frac{D-1}{2}}\frac{w_{2l}^{(a)}(\nu
)}{u^{2l}}\right] \\
&  + \left. 4\sum_{l=1}^{\frac{D-1}{2}} \frac{lw_{2l}^{(a)}(\nu
)}{D-2l} \right\} .
\end{split}
\end{equation}
The expression on the right of this formula can be easily
evaluated numerically. Note that the energy per unit physical
volume on the brane is determined as $(k_Dz_a)^{D}E^{(a)}$ and
does not depend on the brane position.

Unlike to the single brane parts, the interference term $\Delta E$
in Eq. (\ref{totendec}) is finite and is determined by the
formula:
\begin{equation}
\begin{split} \label{DeltaE}
\Delta E &= -\frac{\beta _{D+1}}{D} \int_{0}^{\infty }du \,
u^{D}\frac{d}{du} \ln \left| 1-\frac{\bar I_{\nu }^{(a)}(uz_a)\bar
K_{\nu }^{(b)}(uz_b)} {\bar I_{\nu }^{(b)}(uz_b)\bar K_{\nu
}^{(a)}(uz_a)}\right| \\
&= \beta _{D+1} \int_{0}^{\infty }du \, u^{D-1} \ln \left|
1-\frac{\bar I_{\nu }^{(a)}(uz_a)\bar K_{\nu }^{(b)}(uz_b)} {\bar
I_{\nu }^{(b)}(uz_b)\bar K_{\nu }^{(a)}(uz_a)}\right| .
\end{split}
\end{equation}
Note that this part of the vacuum energy is not affected by finite
renormalizations. Now let us check that this quantity obeys the
standard energy balance equation. We expect that in the presence
of the surface energy this equation will be in the form
\begin{equation}\label{enbalance}
dE=-pdV+\sum_{j=a,b}\varepsilon ^{{\mathrm{(surf)}}}_j dS^{(j)},
\end{equation}
where $V$ is the $(D+1)$-volume in the bulk and $S^{(j)}$ is the
$D$-volume on the brane $y=j$ per unit coordinate volume on the
brane,
\begin{equation}\label{VSj}
V=\int_{a}^{b}dy\, e^{-Dk_Dy}, \quad S^{(j)}=e^{-Dk_Dj},\quad
j=a,b,
\end{equation}
In Eq. (\ref{enbalance}), $p$ is the perpendicular vacuum stress
on the brane and is determined by the vacuum expectation value of
the ${}^{D}_{D}$-component of the bulk energy-momentum tensor:
$p=-\langle 0| T_{D}^{{\mathrm{(vol)}}D}| 0 \rangle $. Combining
equations (\ref{enbalance}) and (\ref{VSj}), one obtains
\begin{equation}\label{dEdzj}
\frac{\partial E}{\partial
z_j}=\frac{n^{(j)}p^{(j)}-Dk_D\varepsilon
^{{\mathrm{(surf)}}(j)}}{(k_Dz_j)^{D+1}},
\end{equation}
with $p^{(j)}$ being the perpendicular vacuum stress on the brane
at $y=j$. In Ref. \cite{Saha03} it has been shown that these
stresses can be presented as sums of the self-action and
interaction parts:
\begin{equation}
p^{(j)}=p^{(j)}_1+p^{(j)}_{{\mathrm{(int)}}},\quad j=a,b.
\label{pintdef}
\end{equation}
The first term on the right is the pressure for a single brane at
$y=j$ when the second brane is absent, and
$p^{(j)}_{{\mathrm{(int)}}}$ is induced by the presence of the
second brane. The latter determines the interaction forces between
the branes and is defined by the formula \cite{Saha03}
\begin{equation}
\begin{split}
p^{(j)}_{{\mathrm{(int)}}}= & -n^{(j)}k_D^{D+1}z_j^{D+1}\beta
_{D+1} \int _{0}^{\infty }du\, u^{D-1} \left[ 1+DB_j\frac{2\zeta
B_j+(1-4\zeta )\tilde A_j}{B_j^2(u^2z_j^2+
\nu ^2)-A_j^2}\right] \\
& \times \, \frac{\partial }{\partial z_j}\ln \left| 1-\frac{\bar
I_{\nu }^{(a)}(uz_a)\bar K_{\nu }^{(b)}(uz_b)}{\bar I_{\nu
}^{(b)}(uz_b)\bar K_{\nu }^{(a)}(uz_a)}\right|  . \label{pint1}
\end{split}
\end{equation}
Assuming that relation (\ref{dEdzj}) is satisfied for the single
brane parts, for the interference parts one obtains
\begin{equation}\label{dDeltaEdzj}
\frac{\partial \Delta E}{\partial
z_j}=\frac{n^{(j)}p_{{\mathrm{(int)}}}^{(j)}-Dk_D\Delta
\varepsilon ^{{\mathrm{(surf)}}}_j}{(k_Dz_j)^{D+1}}.
\end{equation}
Now by taking into account expressions (\ref{emt2pl4}),
(\ref{DeltaE}), (\ref{pint1}) for the separate terms in this
formula, we see that this relation indeed takes place. Hence, we
have checked that the vacuum energies and effective pressures on
the branes obey the standard energy balance equation. Note that
here the role of the surface energy is crucial and the vacuum
forces acting on the brane [determined by
$p^{(j)}_{{\mathrm{(int)}}}$], in general, can not be evaluated by
a simple differentiation of the total vacuum energy.

\section{Conclusion}

\label{sec:Conc}

The natural appearance of AdS in a variety of situations has
stimulated considerable interest in the behavior of quantum fields
propagating in this background. In the present paper we have
investigated the expectation value of the surface energy-momentum
tensor induced by the vacuum fluctuations of a bulk scalar field
with an arbitrary curvature coupling parameter satisfying Robin
boundary conditions on two parallel branes in AdS spacetime. The
Wightman function and the vacuum expectation value of the bulk
energy-momentum tensor for this geometry are investigated in Ref.
\cite{Saha04a} (see also Ref. \cite{Knap03} for the case of the
Randall-Sundrum braneworld). By making use the expression for the
surface energy-momentum tensor from Ref. \cite{Saha03} and the
boundary conditions on the branes, the expectation values of the
surface energy density and stresses are expressed via the
expectation values of the field square on the branes. As an
regularization procedure for the latters we use the generalized
zeta function technique, in combination with contour integral
representations. Using the Cauchy's theorem on residues, we have
constructed an integral representations for the zeta functions on
both branes, which are well suited for the analytic continuation.
These functions are presented as sums of two terms. The first ones
correspond to the zeta functions for single branes when the second
brane is absent. The second terms are induced by the presence of
the second brane and are finite at the physical point. For the
analytic continuation of the single brane zeta functions we
subtract and add to the integrands leading terms of the
corresponding asymptotic expansions, and present them as sums of
two parts. The first one is convergent at the physical point and
can be evaluated numerically. In the second, asymptotic part the
pole contributions are given explicitly. As a consequence, the
single brane surface Casimir energies for separate L- and
R-regions contain pole and finite contributions. The remained pole
term is a characteristic feature for the zeta function
regularization method and has been found in the calculations of
the total Casimir energy for many cases of boundary geometries. As
in the case of the total vacuum energy, the renormalization of
these terms can be performed by using the brane tensions. For an
infinitely thin brane taking L- and R-regions together, in odd
spatial dimensions the pole parts cancel and the surface Casimir
energy is finite. In this case the total surface energy per unit
physical volume on the brane (surface tension) does not depend on
the brane position and can be directly evaluated by making use
formula (\ref{epsLR1}). The results of the corresponding numerical
evaluation for minimally and conformally coupled scalar fields are
presented in Figure \ref{fig1}. As seen from the corresponding
graphs, in dependence of ratio of coefficients in the boundary
condition the surface energy for a single brane can be either
negative or positive. The cancellation of the pole terms coming
from oppositely oriented faces of infinitely thin smooth
boundaries takes place in very many situations encountered in the
literature. It is a consequence of the fact that the second
fundamental forms are equal and opposite on the two faces of each
boundary. In even dimensions there is no such a cancellation.

The surface densities induced on the brane by the presence of the
second brane are investigated in Section \ref{sec:2brane}. The
corresponding contributions to the zeta functions are finite at
the physical point giving finite quantities for all nonzero values
of the interbrane distance and are not affected by finite
renormalizations. The corresponding energy densities are
determined by formula (\ref{emt2pl3}) and are located on the
surfaces corresponding to the region between the branes. They can
also be presented in another form given by Eq. (\ref{emt2pl4}). In
the limit of large AdS radius, $k_D\to 0$, from this formula the
result for two parallel plates on the Minkowski bulk is recovered.
For large distances between the branes, the surface densities
induced by the second brane are exponentially suppressed by the
factor $\exp [k_D (2\nu +D)(a-b)]$ for the brane at $y=a$ and by
the factor $\exp [2\nu k_D (a-b)]$ for the brane at $y=b$. The
exponential suppression also takes place in the large mass limit.
From the viewpoint of an observer living on the brane at $y=j$,
the $D$-dimensional Newton's constant is related to the
higher-dimensional fundamental Newton's constant by formula
(\ref{GDj}) and for large interbrane separations is exponentially
small on the brane $y=b$. The corresponding effective cosmological
constant generated by the second brane  is determined by Eq.
(\ref{effCC}) and is suppressed to compared with the corresponding
Planck scale quantity in the brane universe by the factor $\exp
[k_D (2\nu +D)(a-b)]$, assuming that the AdS inverse radius and
the fundamental Planck mass are of the same order. In the original
Randall-Sundrum model with $D=4$, for a scalar field with the mass
$|m^2|\lesssim k_D^2$, and interbrane distances solving the
hierarchy problem, the value of the cosmological constant on the
visible brane by order of magnitude is in agreement with the value
suggested by current cosmological observations without an
additional fine tuning of the parameters.

The vacuum energy localized on the boundary plays an important
role in the consideration of the energy balance. In Section
\ref{sec:enbal} we consider the total vacuum energy in the region
between the branes, evaluated as a sum of zero-point energies for
elementary oscillators. It is argued that this energy differs from
the energy, obtained by the integration of the bulk energy density
over the region between the branes. We show that this difference
is due to the presence of the surface energy located on the
branes. Further we briefly outline the procedure for the
regularization of the total vacuum energy by the zeta function
technique. This energy is presented as a sum of single branes and
interference parts. The single brane terms contain pole and finite
contributions and explicit formulae are given for both these
parts. In the calculations of the total vacuum energy for a single
brane in odd spatial dimensions the pole parts of the energies,
coming from L- and R-regions, cancel out and a finite result
emerges. This energy per unit physical volume on the brane is
independent of the brane position. For the geometry of two
parallel branes, the interference part in the total vacuum energy
is given by formula (\ref{DeltaE}) and is finite for all nonzero
values of the interbrane separation. Further, we have shown that
the induced vacuum densities and vacuum effective pressures on the
branes satisfy the energy balance equation (\ref{enbalance}) with
the inclusion of the surface terms, which can also be written in
the form (\ref{dDeltaEdzj}).

\section*{Acknowledgments}

The author acknowledges the hospitality of the Abdus Salam
International Centre for Theoretical Physics, Trieste, Italy. This
work was supported in part by the Armenian Ministry of Education
and Science, Grant No. 0887.

\end{document}